\documentclass[%
 reprint,
 amsmath,amssymb,
 aps,
 pre
]{revtex4-1}

\usepackage{graphicx}
\usepackage{dcolumn}
\usepackage{bm}
\usepackage{hyperref}
\usepackage{amssymb}
\usepackage{amsmath}
\usepackage{float}
\usepackage{tcolorbox}

\tcbuselibrary{breakable}
\tcbset{
  width=\columnwidth,
  halign=justify,
  center,
  breakable,
}

\newcommand{\A}{\bm{A}}
\newcommand{\bb}{\bm{b}}

\newcommand{\bt}{\boldsymbol{\theta}}
\newcommand{\MCH}{\mathcal{H}}
\def\multiset#1#2{\ensuremath{\left(\kern-.3em\left(\genfrac{}{}{0pt}{}{#1}{#2}\right)\kern-.3em\right)}}

\begin{document}


\title{Urban Boundary Delineation from Commuting Data with Bayesian Stochastic Blockmodeling: Scale, Contiguity, and Hierarchy}

\author{Sebastian Morel-Balbi}%
\email{smorel@hku.hk}%
\affiliation{%
Institute of Data Science, University of Hong Kong, Hong Kong, China
}%

\author{Alec Kirkley}%
 \email{alec.w.kirkley@gmail.com}
\affiliation{%
 Institute of Data Science, University of Hong Kong, Hong Kong, China\\
 Department of Urban Planning and Design, University of Hong Kong, Hong Kong, China \\
 Urban Systems Institute, University of Hong Kong, Hong Kong, China
}%

\begin{abstract}
A common method for delineating urban and suburban boundaries is to identify clusters of spatial units that are highly interconnected in a network of commuting flows, each cluster signaling a cohesive economic submarket. It is critical that the clustering methods employed for this task are principled and free of unnecessary tunable parameters to avoid unwanted inductive biases while remaining scalable for high resolution mobility networks. Here we systematically assess the benefits and limitations of a wide array of Stochastic Block Models (SBMs)---a family of principled, nonparametric models for identifying clusters in networks---for delineating urban spatial boundaries with commuting data. We find that the data compression capability and relative performance of different SBM variants heavily depends on the spatial extent of the commuting network, its aggregation scale, and the method used for weighting network edges. We also construct a new measure to assess the degree to which community detection algorithms find spatially contiguous partitions, finding that traditional SBMs may produce substantial spatial discontiguities that make them challenging to use in general for urban boundary delineation. We propose a fast nonparametric regionalization algorithm that can alleviate this issue, achieving data compression close to that of unconstrained SBM models while ensuring spatial contiguity, benefiting from a deterministic optimization procedure, and being generalizable to a wide range of community detection objective functions.

\end{abstract}

\maketitle

\section{\label{sec:intro} Introduction}

Urban boundary delineation is a common task in which geospatial data are analyzed to determine disjoint regions of cohesive socioeconomic activity suitable for administrative redistricting, policy interventions, and data smoothing among other applications~\cite{openshaw1995algorithms,moreno2021metropolitan,morrill1999metropolitan,openshaw1995classifying}. This analysis may be performed at various spatial scales, depending on whether a single urban boundary (such as a Metropolitan Statistical Area \cite{us2020omb}) or a collection of boundaries at the sub-city scale \cite{spielman2013using} is required. The precise means by which we determine urban boundaries has been a central problem for geographers and social scientists for over a cenury~\cite{sauer1918geography, dickinson2013city}, as it is well known that small variations in these boundaries can have significant effects on resource allocation, democratic engagement, and the measurement of economic and environmental impact \cite{uchiyama2017methods,arcaute2015constructing,durst2018racial,durst2023gerrymandering}. 

Many different methods have been proposed to construct urban boundaries for cities worldwide \cite{mortoja2020most,thomas2018city,rozenblat2020extending}. Generally speaking, these delineation methods have the shared goal of identifying a partition of the underlying geographic area into clusters such that spatial units that fall within the same cluster are in some way similar to each other and dissimilar to the rest of the clusters. Commuting flow data, which tabulates how many people live in a particular region and commute to work in another region (which may be the same), is critical for capturing unified labor markets across space, as the economic cohesivity of an area is related to workers' willingness to commute within the area \cite{harris1945nature,berry1960impact, berry1968metropolitan}. Consequently, commuting flows from census survey data have been widely used to delineate urban statistical areas and ``megaregions'' for decades, particularly in the United States \cite{adams1999metropolitan,us2020omb,dewar2007planning, innes2010strategies, gottmann1964megalopolis, nelson2016economic}. The precise definition of these boundaries can have an impact on a range of statistical analyses and federal funding decisions \cite{bickers2004interlocal,brazil2022investing}, so it is crucial that a sound statistical methodology is used to construct the urban boundaries from the commuting flows. 

A common representation of commuting flows for boundary delineation is as a network in which nodes are spatial regions and an edge going from one region to another indicates commuting flow between the regions \cite{barthelemy2016structure}. To indicate the amount of commuting flow between a pair of nodes, these networks can either possess edge weights or allow for multiple unweighted edges between a single node pair (i.e. the network is a \emph{multigraph}). Various methods have been proposed in the complex networks and urban science literature in recent years to uncover cohesive urban regions and subregions from the network representation of commuting flows \cite{patuelli2007network,schleith2018assessing,shen2019delineating,de2010functional}. These methods rely on the idea of clustering the nodes to find highly connected groups---known as \emph{communities} in the network science literature~\cite{fortunato2010community}---of spatial units such that there is sparse connectivity between communities. In the context of complex networks, this task is known as \emph{community detection}. 

Modularity maximization~\cite{newman2004finding} is arguably the most popular method for urban boundary delineation with community detection~\cite{nelson2016economic,verhetsel2018assessing,shen2019delineating,kirkley2020information,yu2020mobile,aparicio2023spatiotemporal}. However, this method is known to lead to misleading partitions even for non-spatial networks~\cite{fortunato2010community, guimera2004modularity, fortunato2007resolution, good2010performance, peixoto2023implicit}. Modularity maximization can infer high-scoring partitions even in completely random graphs that by definition have no community structure~\cite{guimera2004modularity}, and suffers from a well known ``resolution limit'' in which highly connected communities of nodes smaller than a certain size cannot be detected \cite{fortunato2007resolution}. 

Considerable effort has been put into developing robust and principled methods to prevent overfitting for community detection tasks in non-spatial networks. The canonical models to achieve this kind of inference are Stochastic Block Models (SBMs)~\cite{holland1983stochastic, wasserman1987stochastic, karrer2011stochastic, peixoto2019bayesian, peixoto2017nonparametric, vaca2022systematic}, which have proven to be extremely useful in extracting mesoscale information in complex networks while remaining robust to overfitting. These methods have the advantage of being completely nonparametric~\cite{peixoto2017nonparametric}, avoiding the choice of the number of clusters or a measure of dissimilarity between clusters, both of which may have undesirable effects on the size and shape of spatial boundaries~\cite{kirkley2022spatial}. This model parsimony is achieved by utilizing the Minimum Description Length (MDL) principle from information theory \cite{grunwald2007minimum}, which states that the best model for a set of data is the one that permits its shortest description (i.e. best compression) in terms of bits of information. Due to its flexible approach for constructing parsimonious nonparametric models of discrete data, the MDL approach has seen use in a broad array of network inference tasks within and beyond community detection \cite{peixoto2019bayesian,hric2016network,kirkley2024identifying,peixoto2017modelling,kirkley2023compressing,kirkley2024dynamic}. With model parsimony in mind, Stochastic Blockmodels have become an increasingly popular method for urban boundary delineation in recent years \cite{he2020demarcating,wang2023mesoscale,zhang2020measuring,zhang2022structural}.

Unfortunately, as is the case for community detection methods in general, SBMs are not explicitly designed to perform \emph{regionalization}---spatially contiguous clustering---which is typically what is required for urban boundary delineation. Regionalization methods can be broadly categorized into implicit or explicit models according to how they deal with the spatial constraints~\cite{duque2007supervised, aydin2021quantitative}. Implicit models generally use non-spatial clustering methods to produce an initial clustering whose spatial contiguity is not guaranteed. From this initial clustering, subsequent post-processing steps are applied to obtain the contiguous partitions~\cite{openshaw1973regionalisation, openshaw1995classifying}. Explicit models, on the other hand, incorporate the constraints directly into the model, thereby ensuring spatial contiguity~\cite{duque2007supervised}. While explicit models are generally preferred because of the rigor with which they apply spatial constraints, both have found widespread use, and both have a set of common characteristics and limitations. 

As they are by nature implicit algorithms for regionalization tasks, community detection methods require considerable post-processing and parameter tuning to enforce spatial cluster contiguity for urban boundary delineation tasks~\cite{nelson2016economic}. These post-processing and parameter tuning choices can have a considerable effect on the spatial partitions learnable from the algorithm, potentially forcing some communities to have high levels of heterogeneity or to be split into multiple smaller communities with no statistical justification \cite{peixoto2019bayesian}. Furthermore, there are a wide range of SBM models available for inference---including methods that capture hierarchical community structure \cite{peixoto2014hierarchical}, ordered community structure \cite{peixoto2022ordered}, community structure in networks with weighted edges, or community structure in multigraphs where multiple edges may run between each pair of nodes \cite{peixoto2019bayesian}---and it is unclear which method should be preferred for the task of urban boundary delineation. There are consequently a few key questions that are important to address when considering the use of SBMs for urban boundary delineation:
\begin{enumerate}
    \item Which commuting network representation (e.g. weighted network or multigraph) and SBM model (e.g. standard, hierarchical, or ordered) should be used for a given urban boundary delineation application?
    \item How does the spatial extent of the commuting dataset and the scale of data aggregation affect the results from these algorithms?
    \item How can we use SBMs to perform urban boundary delineation without requiring extensive post-processing of the inferred communities to achieve spatial contiguity?
\end{enumerate}

We aim to address these three questions in this paper through two major contributions: (1) We systematically investigate the performance and limitations of a wide array of SBM models for urban boundary delineation with commuting data at multiple spatial scales; (2) We extend the SBM to explictly extract contiguous urban boundaries by optimization with a fast greedy agglomerative algorithm and compare its results with these SBM variants. Based on our results, we provide the following practical recommendations:
\begin{tcolorbox}[title=Recommendations for urban boundary delineation with Stochastic Blockmodels]
  \begin{enumerate}

      \item The SBM description length \cite{peixoto2019bayesian} and the contiguity violation measure (CVM) we propose in Section~\ref{sec:results} can be used as guidelines to assess the ``goodness'' of the regionalization obtained by different SBM models.
  
      \item Generally, one should use a weighted network representation rather than a multigraph representation when fitting networks of commute flows to SBMs, as this provides better compression according to the associated SBMs.
      
      \item If spatial contiguity or speed is required, use the fast greedy agglomeration method proposed in this paper. If maximum data compression is desired, use hierarchical block models \cite{peixoto2014hierarchical}, of which the weighted, nested SBM is a good overall choice.

      \item If parsimony in the number of inferred groups is important, use the fast greedy method of this paper. Alternatively, use a nested SBM and inspect the layers of the hierarchy.

      \item If the data aggregation scale is small (e.g. census tracts), use the fast greedy method of this paper to avoid substantial cluster discontiguities and post-processing. If the aggregation scale is large, SBMs find more well-connected partitions and can be considered. 
      
  \end{enumerate}
\end{tcolorbox}

The paper is organized as follows. In Section~\ref{sec:methods} we introduce the techniques and tools used throughout this paper, namely Stochastic Block Models (SBMs) and the Minimum Description Length (MDL) principle. We also describe the objective functions used to infer urban boundaries using SBMs and construct a new greedy agglomerative procedure to minimize them. In Section~\ref{sec:results} we describe the datasets used in this work and present our results from fitting a broad class of SBM models and the proposed fast greedy algorithm to a wide range of metropolitan areas across the continental U.S. at different scales of spatial aggregation. The paper is finalized in Section~\ref{sec:conclusions} with a conclusion. 

A software implementation of the greedy agglomerative algorithm described in this paper is available at \url{https://github.com/seb310/urban_boundary_delineation}.

\section{\label{sec:methods} Methods}

We represent the commuting flows within a given spatial area as a network with $N$ nodes representing smaller spatial units and  
$E$ edges representing the flows between these units. As the commute flows between spatial units need not be reciprocated, we model the commute flow networks as directed networks. It is often the case that there is more than one commute between the same two spatial units, meaning there are two naturally arising network representations we can leverage. In the first representation, we consider each commute as an individual edge of weight $1$. By proceeding this way, we generate \emph{multigraphs}, i.e., networks where multiple connections between nodes are allowed. Multigraphs can be described by an \emph{adjacency matrix} $\bm{A}$ of the form
\begin{equation} \label{eq:adjacency_matrix}
    A = \begin{pmatrix} 
    w_{11} & \dots  & w_{1N}\\
    \vdots & \ddots & \vdots\\
    w_{N1} & \dots  & w_{NN} 
    \end{pmatrix},
\end{equation}
where $w_{ij}$ represents the number of edges (commutes) that run from node $i$ to node $j$. An alternative representation is to consider \emph{weighted} networks so that rather than assigning an individual edge for each commute from $i$ to $j$, we connect the two nodes with a single directed edge and then \emph{weigh} it by annotating it with the total commute flow between the two nodes. Mathematically, weighted networks can be represented by an adjacency matrix whose elements are the weights of the corresponding edges. In the specific case of commute flows, these weights will be integers---the total number of commutes from $i$ to $j$---and the adjacency matrix for the weighted representation of the network of commute flows will still be given by Eq.~\ref{eq:adjacency_matrix}. While the weighted network and multigraph representations are equivalent in this respect, they differ substantially in the way we model the edges statistically, as we will discuss.

\subsection{Stochastic Block Model Inference and the MDL Principle}
\label{sec:SBM-MDL}

As mentioned in Section~\ref{sec:intro}, given a network of commute flows, the urban boundary delineation task is equivalent to that of inferring communities of nodes in the network based solely on the graph's connectivity, and a principled family of methods to perform this kind of task are known as Stochastic Block Models (SBMs)~\cite{holland1983stochastic, anderson1992building, karrer2011stochastic, peixoto2019bayesian}. In its simplest formulation, an SBM partitions a set of $N$ nodes into $B$ groups (typically with $B \ll N$) so that each node $i\in \{1,\ldots,N\}$ is assigned a group label $b_i \in \{1, \ldots, B\}$. Then, edges are placed at random among the nodes such that the probability $p_{b_ib_j}$ that an edge is placed from node $i$ to node $j$ depends only on the group memberships of $i$ and $j$.  

Throughout this work, we will consider a variation of the SBM in which the number of edges $e_{rs}$ between groups $r$ and $s$ is fixed exactly (rather than defined only in expectation), and where edges are distributed uniformly among nodes while fixing the counts $e_{rs}$ for all community pairs $r,s$. This variation is known as the \emph{microcanonical} stochastic block model and has been studied extensively~\cite{peixoto2012entropy, peixoto2017nonparametric, peixoto2014hierarchical, peixoto2013parsimonious}. Microcanonical SBMs can be extended to account for a wide variety of different networks, such as multigraphs~\cite{peixoto2012entropy, peixoto2019bayesian}, directed networks~\cite{peixoto2012entropy, peixoto2017nonparametric}, weighted networks~\cite{peixoto2018nonparametric}, hierarchical block models~\cite{peixoto2014hierarchical, peixoto2017nonparametric}, networks whose nodes are annotated with metadata~\cite{newman2016structure, hric2016network}, and networks with ordered community structure~\cite{peixoto2022ordered}. In this study, we focus our attention on the following (microcanonical) variations:
\begin{enumerate}

    \item The standard stochastic blockmodel, which draws edges uniformly at random while preserving only the number of edges $e_{rs}$ flowing between nodes with each community pair $r,s$.

    \item The degree-corrected stochastic block model (DC-SBM)~\cite{karrer2011stochastic}, which accounts for the heterogeneous degree distributions commonly observed in real-world networks by also preserving the connection strength of each node.

    \item The nested (hierarchical) stochastic block model (nSBM)~\cite{peixoto2014hierarchical, peixoto2017nonparametric}, which constructs a network hierarchy allowing the extraction of communities at multiple scales.

    \item The ranked stochastic block model (rSBM)~\cite{peixoto2022ordered}, where groups are ordered according to some latent ranking and edges flow preferentially downstream (or upstream) according to this ordering.
    
\end{enumerate}
We consider all possible combinations of these models in both the multigraph and weighted network settings, and will discuss the precise formulation of the simplest variant of the SBM from the information theoretic viewpoint in Sec.~\ref{sec:greedy}.

A fundamental characteristic that makes SBMs highly attractive models is their generative nature, which allows for principled inferential approaches based around Bayesian techniques~\cite{wasserman1987stochastic, peixoto2018nonparametric, peixoto2013parsimonious, peixoto2015inferring}. Given any choice of an SBM flavor, it will generate networks with probabilities
\begin{equation}
    P(\A | \bt, \bb),
\end{equation}
where $\A$ is the adjacency matrix, $\bb$ is the node partition, and $\bt$ are other model parameters. Network inference, or a posteriori blockmodelling, is the task of fitting SBM models to data in an attempt to uncover the most likely partition $\bb$ that gave rise to an observed network $\A$ \footnote{Strictly speaking, from a Bayesian perspective, the aim is to determine the posterior distribution and not necessarily the most likely network partition.}.
This can be achieved by appealing to Bayes' rule
\begin{equation} \label{eq:bayes}
    P(\bb | \A) = \frac{P(\A | \bb)P(\bb)}{P(\A)},
\end{equation}
where $P(\bb | \A)$ is the posterior distribution,
\begin{align}
P(\A | \bb) = \int  P(\A | \bt, \bb) P(\bt | \bb)d\bt    
\end{align}
is the marginal likelihood that integrates over any model parameters $\bm{\theta}$, $P(\A)$ is the evidence that acts as a normalizing constant, and $P(\bt | \bb)$ and $P(\bb)$ are the prior distributions encoding our beliefs about the network partition and the mixing preferences of nodes in different groups before we have observed any data. The computation of the marginal likelihood $P(\A | \bb)$ can in general be challenging, but it is trivial for the microcanonical models with hard constraints that we will be discussing. 

By maximizing the left-hand side of Eq.~\ref{eq:bayes}, we can infer the most likely partition $\hat\bb$---called the \emph{Maximum A Posteriori} estimator for the communities---that explains the observed network $\A$ under the stochastic block model. This is a highly non-convex optimization task which is most commonly tackled using Markov chain Monte Carlo (MCMC) methods \cite{peixoto2020merge} that sample the space of partitions $\bm{b}$ stochastically. 

For hierarchical (nested) SBM variants \cite{peixoto2014hierarchical}, the partition of interest is a hierarchical partition $\bm{b}=\{\bm{b}^{(l)}\}_{l=1}^{L}$ of the nodes. In other words, the partition $b^{(l)}$ at level $l$ in the hierarchy is a refinement of the partition at level $l-1$, so that groups at level $l$ are merged to get groups at level $l-1$. Hierarchical SBMs are more challenging to optimize but return a set of multiple nested partitons which can be chosen from in any given application. They have been shown to resolve the resolution limit \cite{fortunato2007resolution} at their lowest hierarchical level. We explore both hierarchical and non-hierarchical SBMs in Sec.~\ref{sec:results}.

A desirable feature of Bayesian approaches is that they are robust against overfitting, which can be observed by making a connection with information theory. We can re-write the numerator on the right-hand side of Eq.~\ref{eq:bayes} as
\begin{equation} \label{eq:mdl}
    P(\A | \bb) P(\bb) = 2^{-\Sigma(\A, \bb)},
\end{equation}
where $\Sigma(\A, \bb)$ is known as the \emph{description length}~\cite{grunwald2007minimum} of the network and corresponds to the negative logarithm (generally taken base 2) of the joint distribution of the network and the model,
\begin{equation} \label{eq:dl}
    \Sigma(\A, \bb) = - \log_2 P(\bb) - \log_2 P(\A | \bb).
\end{equation}
It is a well-known fact from information theory that if a random variable $x$ occurs with probability $P(x)$, then the number of bits needed to transmit the outcome to a receiver using an optimal encoding is lower bounded by $-\log_2 P(x)$~\cite{mackay2003information}. Therefore, the first term in Eq.~\ref{eq:dl} describes the number of bits needed to encode the parameters $\bm{b}$ of the model, while the second term describes the number of bits required to encode the network $\bm{A}$ once the model parameters $\bm{b}$ are known. Since the evidence in Eq.~\ref{eq:bayes} does not depend on the group partition, maximizing the posterior distribution is equivalent to minimizing the description length. Therefore, the network partition $\bm{b}$ that maximizes the posterior $P(\bm{b}\vert \bm{A})$ corresponds to the choice of model parameters that most compresses the data according to the minimum description length (MDL) principle~\cite{grunwald2007minimum, mackay2003information}.

Using this insight we can equivalently frame the microcanonical SBM inference problem as minimizing the description length of an information transmission process, without resorting to any explicit Bayesian arguments. If we were to transmit the structure of a network $\A$ to a receiver, we could split our message into two parts. In the first part of the message, we transmit the parameters of the model $\bb$; then, we use the second part of the message to transmit the network $\A$. The receiver in turn would first decode the model parameters $\bb$ and then decode the network structure $\A$, which will be constrained based on the information previously transmitted in $\bm{b}$. The description length is a measure of the information needed to encode the network along with the model parameters. We use this approach when describing our greedy agglomerative algorithm. 

We can directly see from Eq.~\ref{eq:dl} why Bayesian SBMs are robust against overfitting. As the complexity of the model increases (by increasing the number of groups), it will better accommodate the data, and $-\log_2 P(\A | \bb)$ will decrease. However, as the complexity of the model increases, so will the number of bits required to transmit it, and $-\log_2 P(\bb)$ will increase. The first term in Eq.~\ref{eq:dl} then acts as a penalty, preventing models from becoming overly complex, and the optimal choice corresponds to a balance between the two terms.

The MDL principle also gives us a principled approach to compare results obtained by different models. Suppose we have two different SBM models, which we call $\MCH_1$ and $\MCH_2$, that correspond to two different hypotheses for the generating mechanism that produced some observed network $\A$. Let $\bb_1$ and $\bb_2$ be the corresponding most likely partitions obtained by maximizing their respective posterior distributions. A principled approach to decide which of the two partition/model combinations better represents the data is to compute the posterior odds ratio~\cite{jaynes2003probability}
\begin{equation}
    \frac{P(\bb_1, \MCH_1 | \A)}{P(\bb_2, \MCH_2 | \A)} = 
    \frac{P(\A | \bb_1, \MCH_1)P(\bb_1 | \MCH_1)P(\MCH_1)}{P(\A | \bb_2, \MCH_2)P(\bb_2 | \MCH_2)P(\MCH_2)},
\end{equation}
where a ratio above (below) $1$ indicates that we should favor model $\MCH_1$ ($\MCH_2$) based on posterior probability.
If we assume that the two models are a priori equally likely ahead of observing any data so that $P(\MCH_1) = P(\MCH_2)$, we have that the posterior odds ratio can be written as
\begin{equation}
    \frac{P(\bb_1, \MCH_1 | \A)}{P(\bb_2, \MCH_2 | \A)} = 
    \frac{P(\A | \bb_1, \MCH_1)P(\bb_1 | \MCH_1)}{P(\A | \bb_2, \MCH_2)P(\bb_2 | \MCH_2)} = 2^{\Sigma_2 - \Sigma_1}.
\end{equation}
Therefore, we should generally prefer the model with the lower description length---equivalently, the model that better compresses the data.

\subsection{A greedy agglomerative method for SBM regionalization}
\label{sec:greedy}

Arguably, the most significant limitation that SBMs present for urban boundary delineation is that they have no notion of spatial contiguity---the partition $\bm{b}$ of the network depends only on commuting flows $\bm{A}$ and is not required to be spatially contiguous. (In other words, SBMs are \emph{implicit} regionalization methods rather than \emph{explicit}.) This issue is not unique to SBMs but is a common characteristic of all community detection methods that solely make use of topological information~\cite{nelson2016economic}. To circumvent this issue, we propose a fast greedy regionalization algorithm based on the MDL principle which does not require the user to specify or fine-tune any free parameters, as is often the case with other spatial regionalization techniques~\cite{duque2007supervised, assunccao2006efficient, aydin2021quantitative}. Our method is similar in spirit to the one introduced in~\cite{kirkley2022spatial}. However, in~\cite{kirkley2022spatial}, the regionalization was performed taking into consideration only node attributes, without considering possible interactions between the spatial units such as commute flows. Here, we extend this method to include both node attributes and node interactions as is required for urban boundary delineation. 

The fast agglomerative algorithm we construct for regionalization with SBMs can be applied to any clustering optimization objective $F(\bm{A},\bm{b})$ of the form 
\begin{equation} \label{eq:general_objective}
F(\bm{A},\bm{b}) = C + \sum_{r=1}^{B} g(r) + \sum_{r,s=1}^{B} f(r, s),
\end{equation}
where $C$ is a function of only global network constants, and $g(r)$ and $f(r, s)$ represent functions specifying the cluster-level properties and the cluster-to-cluster interactions respectively with resepect to the input partition $\bm{b}$. The degree-corrected and ordered variants of the SBM (all with directedness, weights or multi-edges if desired) \cite{karrer2011stochastic,peixoto2022ordered}, as well as modularity \cite{clauset2004finding} and Infomap \cite{rosvall2008maps} can all be written in this form, so our regionalization method can be applied quite generally for regionalization using existing community detection methods. In this paper we apply our regionalization algorithm to clustering objectives $F(\bm{A},\bm{b})=\Sigma(\bm{A},\bm{b})$ from Eq.~\ref{eq:dl} for the weighted and multigraph directed SBMs. 

As discussed in Sec.~\ref{sec:SBM-MDL}, the SBM description lengths $\Sigma(\bm{A},\bm{b})$ whose general form is given by Eq.~\ref{eq:dl} can be derived by considering a multi-part transmission process where we first transmit the partition $\bm{b}$ of the network---with information content given by $-\log_2 P(\bm{b})$---then transmit the network $\bm{A}$ using this partition $\bm{b}$---with information content given by $-\log_2 P(\bm{A}\vert \bm{b})$. We assume that the constants $\{N,E,B\}$ are already known by the receiver, as these will add irrelevant constants to the description length obejctive anyway. Given these constraints, the information content \cite{peixoto2019bayesian} to transmit a network partition with a uniform prior over all partitions with the observed given group sizes is 
\begin{align}
-\log P(\bm{b}) &= -\log \left[\frac{1}{{N-1\choose B-1}}\times \frac{1}{{N\choose n_1,...,n_B}}\right]\\
&= \log {N-1\choose B-1} + \log {N\choose n_1,...,n_B}   
\end{align}
bits, where $n_r$ is the number of nodes in group $r$ and we've let $\log \equiv \log_2$ for convenience. This can be derived by computing the probabilities for drawing (1) the $B$ non-zero group sizes $\bm{n}=\{n_r\}$ uniformly from the ${N-1\choose B-1}$ possibilities that must sum to $N$; and (2) the partition uniformly from all ${N\choose n_1,...,n_B}$ unique partitions with sizes $\bm{n}$. This prior is the same for all non-nested models we consider for our regionalization algorithm.

Although $-\log P(\bm{b})$ is the same across SBM models, for multigraphs and weighted networks the second term $-\log P(\bm{A}\vert \bm{b})$ in the description length differs considerably. Both cases require us to transmit the mixing matrix $\bm{e}=\{e_{rs}\}_{r,s=1}^{B}$, which tells us the number of edges that run between each pair of groups $r,s$. There are $\multiset{B^2}{E}$ unique ways to place the $E$ directed edges between pairs of groups $(r,s)$, allowing some group pairs to potentially have no edges between them. Here, $\multiset{a}{b}={a+b-1\choose b}$ is the multiset coefficient counting the number of unique ways to draw $b$ objects from a set of $a$ distinct objects. Therefore, each SBM model will require an entropy of $-\log \multiset{B^2}{E}^{-1}=\log \multiset{B^2}{E}$ to transmit $\bm{e}$. But once $\bm{e}$ is known, each model must transmit a fundamentally different graph structure.

For the multigraph SBM, there may be any number of edges running from node $i$ to node $j$---each edge in this case represents a commuter who commutes from their home at node $i$ to their workplace at node $j$. Given a node partition $\bm{b}$ and a mixing matrix $\bm{e}$ describing the number of edges within and between these communities, there are $\multiset{n_rn_s}{e_{rs}}$ ways to distribute $e_{rs}$ edges among the $n_rn_s$ node pairs such that the source node is in $r$ and the destination node is in $s$, allowing for multiple edges to run from a node $i$ to a node $j$. Therefore, the description length required to transmit the full multigraph representation given the already known constraints $\bm{b}$ is given by \cite{peixoto2012entropy}
\begin{align}
-\log P_{MG}(\bm{A}\vert \bm{b}) = \log \multiset{B^2}{E} + \sum_{r,s=1}^{B}\log \multiset{n_rn_s}{e_{rs}}.    
\end{align}

In contrast to the multigraph SBM, for the weighted SBM we must transmit a simple graph with at most one edge running between each node pair $(i,j)$ and then transmit the value of the positive integer-valued weight on each edge in multiple steps. Only allowing at most one edge to run from a node $i$ to a node $j$, there are ${n_rn_s \choose e_{rs}}$ ways to distribute $e_{rs}$ edges among the $n_rn_s$ node pairs such that the source node is in $r$ and the destination node is in $s$. (This is in contrast to the multigraph case where we allow for multiple edges between each node pair.) Transmitting the unweighted, simple graph therefore incurs a cost of
\begin{align}
\sum_{r,s=1}^{B}\log {n_rn_s \choose e_{rs}}    
\end{align}
bits. After the graph structure is known, we can transmit the weights on the edges, which is done in a similar manner to the edges by specifying a mixing matrix $\bm{w}=\{w_{rs}\}$ for the weights, where $w_{rs}$ is the total edge weight going from nodes in group $r$ to nodes in group $s$. If we assume that the total weight $W$ among all edges is known---this incurs a negligible information cost as with the other constants---then the amount of information to transmit the weight mixing matrix $\bm{w}$ is $\log\multiset{B^2}{W}$, analogous to transmitting the edge mixing matrix. Then, to transmit the weights on the edges, we can specify how $w_{rs}$ is distributed aming the $e_{rs}$ edges running between the groups. There are ${w_{rs}-1\choose e_{rs}-1}$ ways in which the weight $w_{rs}$ can be distributed among $e_{rs}$ edges while requiring each edge weight to be a positive integer. Putting this altogether and summing over the cluster-level quanitities being transmitted gives the description length for transmitting the weighted network representation of the data given the already known constraints $\bm{b}$, thus
\begin{align}
-\log P_{W}&(\bm{A}\vert \bm{b}) = \log \multiset{B^2}{E} + \log\multiset{B^2}{W} \\
&+ \sum_{r,s=1}^{B}\left[\log {n_rn_s \choose e_{rs}} + \log {w_{rs}-1\choose e_{rs}-1}\right]  \nonumber 
\end{align}

Using Eq.~\ref{eq:dl}, the description lengths for the multigraph and weighted SBMs are then given by%
\begin{align}
\Sigma^{(MG)}(\bm{A},\bm{b}) &= -\log P(\bm{b})-\log P_{MG}(\bm{A}\vert \bm{b}), \label{eq:MGdl}\\
\Sigma^{(W)}(\bm{A},\bm{b}) &= -\log P(\bm{b})-\log P_{W}(\bm{A}\vert \bm{b}). \label{eq:Wdl}
\end{align}
Minimizing the description lengths in Eq.~\ref{eq:MGdl} and Eq.~\ref{eq:Wdl} identifies the MDL-optimal partitions $\bm{\hat b}^{(MG)}$ and $\bm{\hat b}^{(W)}$ of the network under the multigraph and weighted SBM models respectively. We can then see why the choice of network representation may influence the outcome of the clustering procedure: the two description lengths above may compress the data differently (e.g. have different minimum description length values) and have different partitions $\bm{\hat b}$ at which they are minimized. 

To minimize an SBM description length such as Eq.~\ref{eq:MGdl} or Eq.~\ref{eq:Wdl} exactly, one must search over all partitions $\bm{b}$ of the network into $B\leq N$ groups, which quickly becomes computationally intractable. Therefore, one often resorts to stochastic optimization algorithms based on Markov chain Monte Carlo (MCMC) sampling~\cite{peixoto2014efficient, peixoto2020merge}, which are theoretically guaranteed to find the optimal partition $\bm{b}$ under certain conditions \cite{newman1999monte}. MCMC sampling iteratively samples partitions $\bm{b}$ and randomly proposes new partitions $\bm{b}'$ to sample next, which are accepted based on the current partition's description length $\Sigma(\bm{A},\bm{b})$ and the proposed partition's description length $\Sigma(\bm{A},\bm{b}')$. In practice SBM optimization is often implemented using simulated annealing \cite{peixoto_graph-tool_2014}, where moves with $\Sigma(\bm{A},\bm{b}')<\Sigma(\bm{A},\bm{b})$ are always accepted but moves with $\Sigma(\bm{A},\bm{b}')>\Sigma(\bm{A},\bm{b})$ are accepted with a probability that decreases as the simulation progresses according to a ``cooling schedule''. There are two primary practical issues with the standard SBM simulated annealing approach that make it challenging for use in urban boundary delineation tasks: (1) Convergence is often very slow and sometimes requires careful monitoring to determine when to stop the algorithm; (2) The partitions $\bm{b}$ sampled from the algorithm are discontiguous with high probability, and efficient algorithms to sample contiguous partitions of a network often do not exist \cite{najt2019complexity}.  

Here we propose a deterministic greedy regionalization algorithm that preserves the spatial contiguity of the inferred clusters and avoids these disadvantages of MCMC sampling, making it more applicable to the problem of urban boundary delineation. The description lengths in Eq.~\ref{eq:MGdl} and Eq.~\ref{eq:Wdl} can both easily be mapped to the form in Eq.~\ref{eq:general_objective}, so for generality we will consider optimizing any function of this form with our greedy regionalization procedure. The algorithm begins by assigning each node $i$ in the network to its own cluster and progressively merging adjacent clusters by selecting the merge that produces the largest decrease in the objective function until no merge is able to reduce it further. To be able to identify adjacent clusters, we need to introduce the (undirected) network of spatial adjacencies of the geographical area we are considering, which we denote by $G'(N, E')$, where the set of nodes $N$ is the same as in the commute flow network, but the edges $E'$ are now placed between two tracts only if they are spatially adjacent. To clarify, we consider two tracts to be adjacent if they share a common length of border. We can then use this network to inform the merging procedure by only proposing merges between adjacent spatial units. In subsequent steps of the algorithm, we will consider two clusters $r$ and $s$ to be adjacent if there is at least one node $i \in r$ and one node $j \in s$ such that they are connected by a spatial edge $(i,j) \in E'$.

We can then map our network of commute flows on top of the spatial adjacency network and use the spatial adjacency network to inform the merging. It must be noted that the commute flow network will typically contain interactions between nodes/groups that need not be spatially adjacent and must be accounted for during the merging procedure. For any two clusters $r$ and $s$, we can compute the variation in Eq.~\ref{eq:general_objective} caused by their merge as
\begin{align} \label{eq:ddl}
        \Delta &F(r, s) = g((r,s)) - g(r) - g(s) \nonumber \\
        &+ f((r,s),(r,s)) - f(r,s) - f(s,r) - f(r,r) - f(s,s) \nonumber \\
        &+ \sum_{u \in \mathcal{N}^{in}_{(r,s)} \cup \mathcal{N}^{out}_{(r,s)}} [f((r,s),u) + f(u, (r,s))]\\
        &- \sum_{u \in \mathcal{N}^{in}_{(r,s)} \cup \mathcal{N}^{out}_{(r,s)}}[f(r,u)+f(s,u) +f(u,r) + f(u,s)], \nonumber
\end{align}
where $(r,s)$ is the label of the newly merged group and $\mathcal{N}^{in}_{(r,s)} = \mathcal{N}^{in}_r \cup \mathcal{N}^{in}_s$ and $\mathcal{N}^{out}_{(r,s)} = \mathcal{N}^{out}_r \cup \mathcal{N}^{out}_s$ are the sets of in and out-neighbors, in the original commute flow network, of the newly merged group. We have omitted the terms corresponding to the constants $C$ in Eq.~\ref{eq:general_objective}, as these will be identical for all merges and can be computed after the best merge has been selected.

Eq.~\ref{eq:ddl} can be computed in $O(k_r + k_s)$ time, where $k_r$ represents the total degree (in and out) of group $r$. Furthermore, the computed values of Eq.~\ref{eq:ddl} can be stored and referenced in the future without the need to re-evaluate all of them at each step of the algorithm. However, since we now have interactions between nodes that arise from the $f(r,s)$ terms, whenever we merge two clusters $r$ and $s$, we must update all of the stored $\Delta F (u, v)$ values for which $u$ is an in or out-neighbor of the newly formed group $(r,s)$, and $v$ is a \emph{spatially} adjacent neighbor of $u$ other than $(r,s)$. These updates can be performed in $O(k_{(r,s)} \sum_{u \in \mathcal{N}_{(r,s)}} d_u)$ time, where $d_u$ is the number of spatially adjacent clusters to cluster $u$, $\mathcal{N}_{(r,s)} = \mathcal{N}^{in}_{(r,s)} \cup \mathcal{N}^{out}_{(r,s)}$ indicates the entire neighborhood (in and out) of the newly merged group $(r,s)$, and $k_{(r,s)}$ is the total degree (in and out) of the newly merged group. In principle, the value of $d_u$ can be as large as $N-1$. However, from a practical standpoint, these values are typically much smaller, as nodes in a planar graph generally tend to have small degrees, which can, therefore, be considered constant in terms of computational complexity. The update step can then be computed in $O(k_{(r,s)})$ time, and since $k_r\leq 2B-1$ the update procedure is upper bounded by $O(B)$.

The brunt of the computational complexity of our procedure lies in the need to scan through all the $O(B)$ merge pairs to determine the optimal one at each step, as there is no efficient way of keeping track of the best merge pair. The updates make the use of a heap impractical, as many of the entries will change after each merge. This implies that we need to conduct $N$ scans initially, $N-1$ in the next step, and so on, resulting in an overall time complexity for our algorithm of $O(N^2)$. In our experiments we observe a sub-quadratic scaling of the runtimes with $N$ (see Fig.~\ref{fig:exec_time} in Appendix~\ref{appendix:appendix_a}) and that our algorithm is easily scalable to large commuting networks.  This suggests that in practice the various constant-time operations required to maintain the data structures have a higher computational cost than the $O(B)$ scans needed to identify the best merge pair.

One could also apply our method to the modularity objective, in which case we can use heaps for faster optimization since $f(r,s)=0$ and the merges do not affect non-physically adjacent clusters. This would be equivalent to the fast greedy method of \cite{clauset2004finding} but with fewer merge proposals due to the spatial contiguity constraints.

It is important to stress that this agglomerative algorithm is greedy in nature and, as such, offers no guarantee that it will find the optimal spatially contiguous partition. However, we find that it is capable of finding spatially contiguous states with description lengths comparable to those of the more flexible SBM models fitted with MCMC techniques. Additionally, our parameter-free greedy method is arguably easier to implement as it sidesteps many of the challenges often associated with the use of MCMC techniques, such as evaluating the mixing and convergence of the chains or optimizing the tuning parameters.

\begin{figure*}
    \centering
    \includegraphics[width=\textwidth]{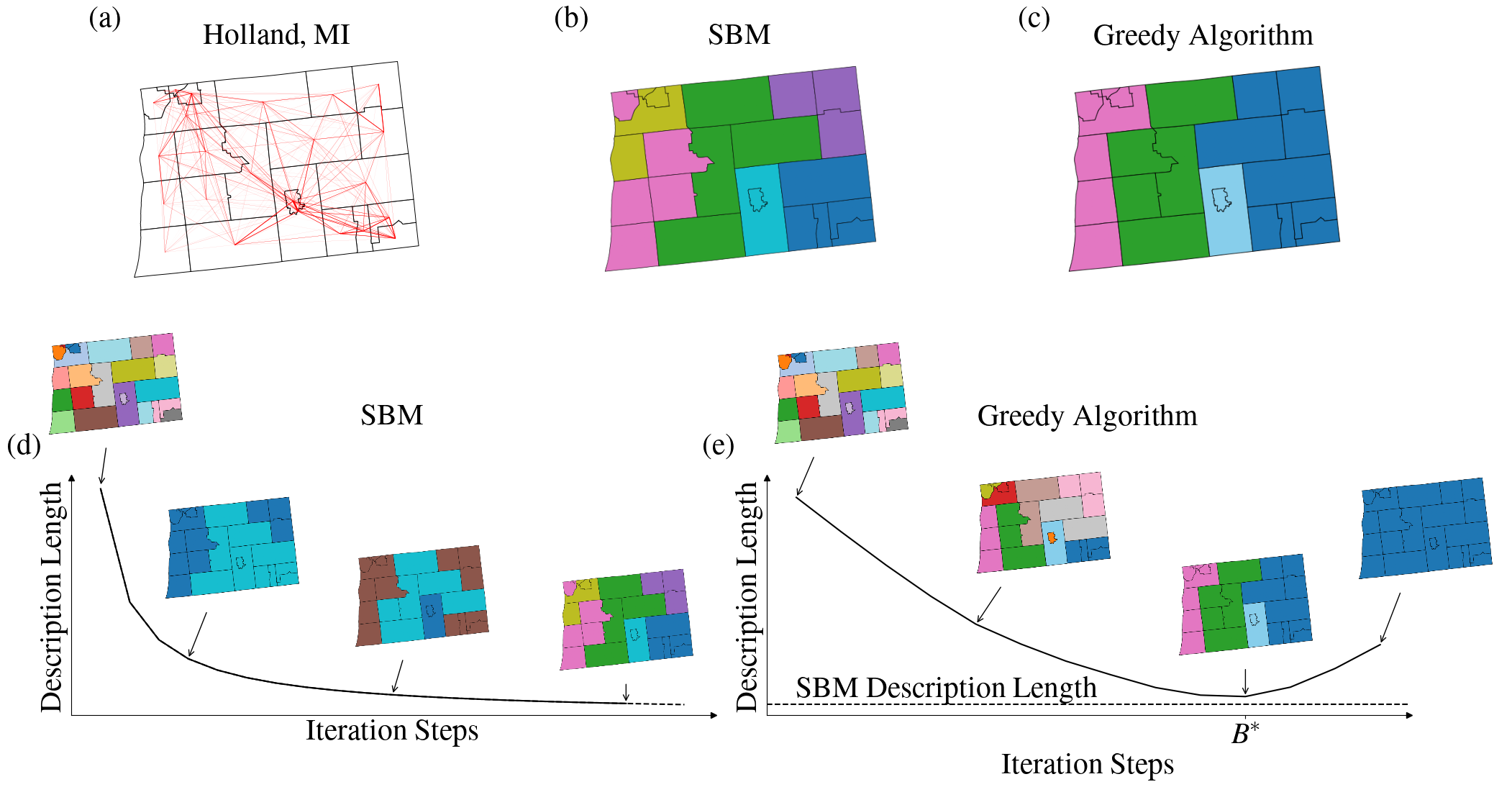}
    \caption{
    Illustration of the methodologies described in the text. (a) The original problem formulation: given a network of commute flows in a metropolitan area, infer a regionalization that captures cohesive commute patterns. (b) Results obtained by minimizing the description length of the weighted SBM (Eq.~\ref{eq:Wdl}) with simulated annealing. (c) Results obtained by minimizing the same description length with the proposed greedy regionalization algorithm described in Sec.~\ref{sec:greedy}. (d) Illustration of the optimization trajectory of simulated annealing. (e) Illustration of the optimization trajectory of the greedy agglomerative method, with the description length obtained by MCMC indicated by the dashed line.
    }
    \label{fig:methodology}
\end{figure*}

Fig.~\ref{fig:methodology} shows a schematic representation of the regionalization procedure for an example network of commuting flows from Holland, Michigan (see Sec.~\ref{sec:results}). Fig.~\ref{fig:methodology}(a) shows an example of the regionalization problem. Given a geographic area composed of a set of fundamental units and a set of commute flows between the units, regionalization aims to partition the region into clusters based solely on the commute flow patterns. Fig.~\ref{fig:methodology}(b) and Fig.~\ref{fig:methodology}(c) show the final clusters obtained by applying the standard weighted SBM and our greedy agglomerative algorithm to the commuting network in panel (a). Although the results are qualitatively similar, the SBM partition is discontiguous, breaking up the pink cluster into two disconnected fragments. Fig~\ref{fig:methodology}(d) and Fig.~\ref{fig:methodology}(e) show the optimization trajectories of the unconstrained SBM with MCMC and regionalization with our greedy procedure respectively. We can see that the SBM MCMC optimizer explores a complex solution space where partitions tend to have disconnected clusters, the description length decreases nearly monotonically, and the algorithm is run until convergence is determined by some heuristic (e.g. the sampler fails to jump to a new partition for some threshold number of move proposals). This may take quite a long time, since it is unclear when to stop the algorithm. Meanwhile, the greedy procedure explores a much simpler space of solutions with contiguous clusters and is run for exactly $N-1$ merge steps, at which point we can inspect all examined partitions and their associated description lengths to find the local optimum for $\bm{\hat b}$.    

As shown in Fig.~\ref{fig:methodology}(e), MCMC methods used to fit SBM models can generally achieve lower description lengths. However, the greedy agglomerative method can generally infer partitions close (in terms of description length) to those inferred by the SBM. Furthermore, as seen in Fig.~\ref{fig:methodology}(b) and Fig.~\ref{fig:methodology}(c), SBM models typically infer discontiguous network partitions, while the greedy approach guarantees contiguity.

\section{\label{sec:results} Results}

\subsection{Data description}

In line with previous studies~\cite{lowe1998patterns,nelson2016economic,he2020demarcating,rae2017big}, we analyze tract-to-tract workplace commuting data obtained from the American Community Survey (ACS)~\cite{acs2010}, which collects, among other things, data on commuting patterns at various geographic levels, including the origin and destination of commuters and the total commute flows between different geographic areas. The ACS generates commute flow data every five years using non-overlapping 5-year estimates. In our work, we have used a dataset compiled in~\cite{nelson2016economic} based on the 2006-2010 ACS Census Tract Flow dataset.

Covering the entirety of the United States, census tracts are designed to be relatively homogeneous units with respect to population characteristics and socioeconomic status. Furthermore, with an average population of about 4,000 people, census tracts are generally big enough to contain large amounts of the workforce population~\cite{nelson2016economic}. Additionally, census tracts are part of a well-defined geographical hierarchy, meaning they can be combined to obtain county subdivisions, allowing us to compare regionalization results at different spatial scales. 

To process the commuting flow dataset, we first mapped the origin and destination tracts to nodes in a network. We then constructed directed network representations of the commute flows across three different categories of metropolitan regions. Core-based statistical areas (CBSAs), combined statistical areas (CSAs), and entire states across the U.S. CBSAs are geographic regions consisting of one or more counties anchored by an urban center of at least 10,000 people. They can be both urban and rural in nature and are generally designed to capture unified patterns of socioeconomic ties. CSAs are obtained by combining two or more adjacent CBSAs that have a high degree of social and economic integration (usually measured in terms of commuting patterns). CSAs provide larger geographic units for evaluating economic and demographic trends and can help identify regional economic hubs and labor markets that extend beyond individual metropolitan or micropolitan areas.

TIGER shapefile data~\cite{tiger2019} was used to compute the tract geometries and commute distances, and additional metadata was mapped to each node, including each tract's FIPS code identifier and geographical position. In the case of weighted networks, the total commute flow between two tracts was also mapped to the corresponding edge. Finally, following~\cite{nelson2016economic}, we filtered out all commutes with a distance $d > 160$km. This filtering allows us to remove the most extreme commutes while retaining over $97\%$ of the data. After pre-processing, we are left with 793 networks representing a range of metropolitan areas across the United States.

As previously mentioned, care should be taken when selecting which network representation to use. Arguably, the most natural network representation is to consider each commute as an individual edge and generate multigraphs. An alternative representation is to consider weighted networks so that rather than assigning an individual edge for each commute from $i$ to $j$, we connect the two nodes with a single directed edge and annotate it with the total commute flow between them. The choice of which representation to use has a profound effect on the outcome of the regionalization procedure, as even though the same adjacency matrix describes both representations, the description length required to transmit the network will differ, as shown in Eq.~\ref{eq:MGdl} and Eq.~\ref{eq:Wdl}. In the next section, we show that weighted models tend to significantly outperform multigraph-based ones and should thus be preferred.

\subsection{Tract-level results}

We begin our analysis by assessing the efficacy of the weighted and multigraph representations in capturing the patterns of commute flows across all metropolitan areas. We construct multigraphs and weighted networks for each of the 793 metro regions in our dataset and fit a variety of SBM models on each via the graph-tool Python package~\cite{peixoto_graph-tool_2014}. Specifically, we have considered all possible combinations of nested and non-nested models, ranked and unranked, and degree and non-degree corrected, yielding eight different SBM models: the standard stochastic block model (SBM), the degree-corrected stochastic block model (DC-SBM), the ranked stochastic block model (rSBM), the ranked degree-corrected stochastic block model (rDC-SBM), the nested stochastic block model (nSBM), the nested degree-corrected stochastic block model (nDC-SBM), the ranked nested stochastic block model (rnSBM), and the ranked nested degree-corrected stochastic block model (rnDC-SBM). The equivalent weighted formulations have been considered for weighted graphs. Additionally, we have run our greedy agglmoerative algorithm (GA) on each multigraph and weighted network.

\begin{figure}[!ht]
    \centering
    \includegraphics[width=\columnwidth]{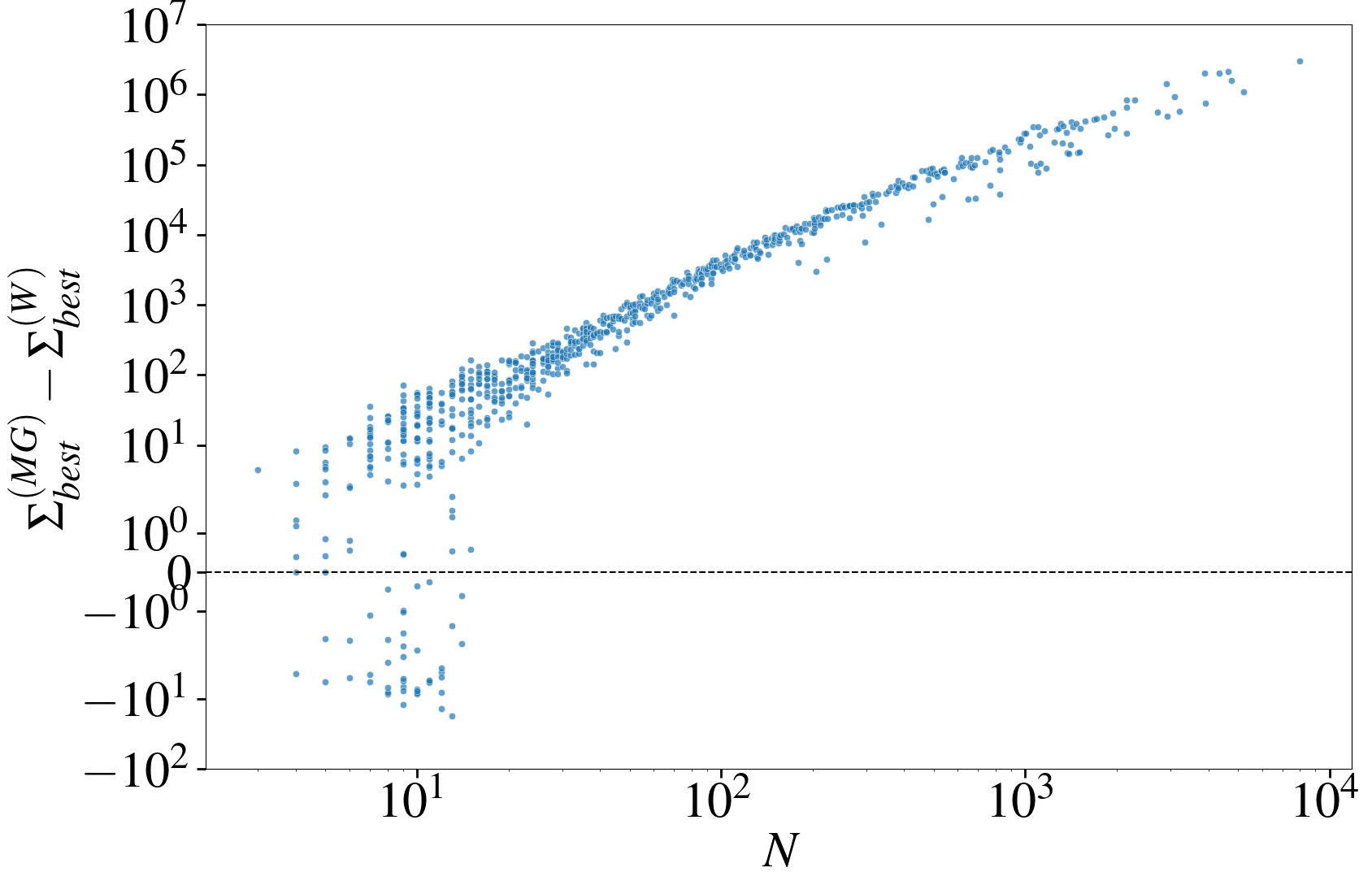}
    \caption{
    Difference in description length between the best-performing multigraph model and the best-performing weighted model across all metropolitan areas as a function of $N$. Positive values of the difference indicate that weighted variants tend to outperform multigraph ones across all metropolitan areas, and the difference in performance grows larger as the number of nodes in the network increases.
    }
    \label{fig:weighted_vs_multigraph}
\end{figure}

Fig.~\ref{fig:weighted_vs_multigraph} shows the difference in description length, $\Sigma^{(MG)}_{best} - \Sigma^{(W)}_{best}$, between the best performing multigraph and weighted models for each metro region in our dataset as a function of the number of nodes $N$ in the network. We can observe that weighted network representations tend to outperform multigraphs across the entire dataset, allowing for better data compression, especially as the networks increase in size \footnote{
    When referring to the \emph{size} of the network, one must note that we are referring to the number of tracts making up the metropolitan region and not to its spatial extent.
    }.
The cases for which multigraph networks provide marginally lower description lengths all have fewer than $20$ nodes. Additionally, multigraph models tend to infer a much larger number of groups than weighted models, making the inferred regionalizations harder to interpret. These findings suggest that weighted models should generally be preferred over multigraph ones when dealing with commute flow data, and we will focus exclusively on them for the remainder of this paper.

\subsubsection{Compression results for weighted networks}

\begin{figure*}[!ht]
    \centering
    \includegraphics[width=\textwidth]{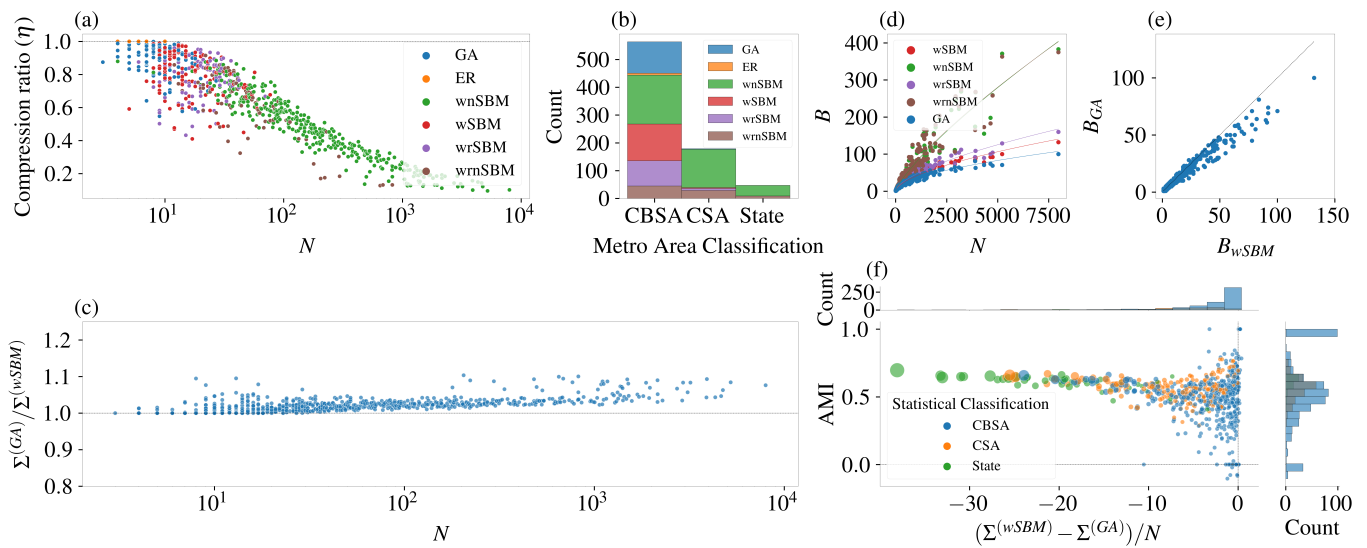}
    \caption{
    (a) Compression ratio for the best performing model as a function of $N$ across all metropolitan areas considered. The dashed horizontal line corresponds to the performance of the null Erdos-Renyi style model (Eq.~\ref{eq:er_dl}). We can observe that as the number of nodes $N$ increases, SBM models achieve increasing compression with respect to the null model. (b) Histogram of the best-performing model across the statistical classifications of the metropolitan areas. While there is more diversity among the best-performing models for CBSAs, as we move from CSA's to States the weighted nested variant of the SBM becomes dominant due to the increasing size $N$. (c) Description length ratio between the regionalizations obtained via the greedy agglomerative method and the best-performing SBM model. While SBMs always outperform the greedy method in terms of compression, the differences are quite modest even for large networks. (d) Number of groups inferred by the models as a function of $N$, with regression lines drawn as a visual guide. Nested models tend to identify more groups than non-nested models. (e) Comparison of the number of groups inferred by the greedy agglomerative method and the standard wSBM, the black dashed line indicating the case $B_{GA} = B_{wSBM}$. Both methods infer a similar number of groups, with the greedy method consistently finding slightly fewer. (f) Adjusted mutual information (AMI) between the partitions inferred with the best-performing SBM model and the greedy algorithm across all metropolitan areas considered as a function of the difference in description length (normalized by network size). The sizes of the points are proportional to $N$.
    }
    \label{fig:compression}
\end{figure*}

To assess the quality of information compression achieved by regionalizing the commuting data, we compare the description lengths under the different models to the description length of the data under a null model that does not exploit cluster structure for data compression. A natural choice is to consider a weighted equivalent of the Erdős–Rényi random graph model (ER model)~\cite{erdHos1960evolution}. We assume once more that the receiver knows the total number of nodes $N$, edges $E$, and the total flow $W$ of commutes over the network. In order to have a null model that can account for both weighted networks and multigraphs, we can consider the following procedure. We first split the total flow across the network into $W$ weights of value $1$ each. Then, we select pairs uniformly at random with repetition. If the selected node pair is not connected, we add an edge and set a weight on that edge equal to 1. If the node pair is already connected, we simply increase the weight on the edge by one. This procedure is exactly the same procedure one would use to generate random directed multigraphs with $N$ nodes and $W$ edges. Since there are $\multiset{N^2}{W}$ ways of distributing $W$ directed multiedges amongst $N$ nodes (allowing for self-edges), the total information cost to transmit a network encoded with this random graph model is
\begin{equation} \label{eq:er_dl}
    \Sigma^{(ER)}(\bm{A}) = \log \multiset{N^2}{W}.
\end{equation}
Equipped Eq.~\ref{eq:er_dl}, we can define a \emph{compression ratio} analogous to the one introduced in~\cite{kirkley2022spatial} that allows us to assess the degree of compression achieved. We define the compression ratio as
\begin{equation}
    \eta(\bm{A}) = \frac{\Sigma^{(model)}(\bm{A},\bm{b})}{\Sigma^{(ER)}(\bm{A})},
\end{equation}
where $\Sigma^{(model)}(\bm{A},\bm{b})$ is the description length for the partition of either a specific weighted SBM model or the greedy algorithm introduced previously (Eq.s~\ref{eq:MGdl}~and~\ref{eq:Wdl}). Values of $\eta$ close to zero indicate that the partitions inferred by the model are efficiently compressing the data. Conversely, a value of $\eta$ equal to one indicates a lack of an underlying group structure so that no partition can achieve better compression than if we had transmitted the entire dataset edge by edge, ignoring any group structure as in the null model. Values of this compression ratio for the best-performing model across all metro regions are shown in Fig.~\ref{fig:compression}(a) as a function of the number of nodes $N$ in the network. For smaller networks, compression ratios tend to be higher, and, in some cases, no model performs better than random. However, there is a considerable spread of $\eta$ values, particularly at low values of $N$, indicating that commute patterns can, in some cases, inform the underlying functional geographic structure even in very small metropolitan areas. As $N$ increases and the networks become increasingly complex, the network can be better compressed by exploiting its clustered structure at multiple scales. In particular, nested SBMs emerge as the dominant models as they capture heterogeneities in commuting patterns at multiple scales simultaneously.

The dominance of nested models is further shown in Fig.~\ref{fig:compression}(b), which shows a histogram of the best-performing models across the three different statistical classifications of metro areas considered in this work. We can observe a high degree of heterogeneity across the best-performing models for smaller metropolitan areas such as CBSAs (which also tend to have fewer tracts). While nested models are still dominant at these scales, high compression levels can, in many cases, be achieved by considering simpler models. However, as the size of the metro region increases and commuting patterns become more complex, nested models consistently outperform all others in terms of the compression ratio achieved. 

Another interesting feature of Figs.~\ref{fig:compression}(a)-(b) is the prevalence, amongst nested models, of the standard nested wSBM as the best-performing model, particularly at large values of $N$. It would appear that some attributes and considerations generally associated with urban regions are not necessary to explain the observed commuting patterns. For example, the relative scarcity of ranked models among the best-performers suggests that, once the clustered structure is accounted for, hierarchical structures often displayed by some urban regions provide redundant or unnecessary information for explaining the commute patterns for a wide fraction of the metropolitan areas considered in this study.

It is important to note though that Fig.~\ref{fig:compression}(a) and Fig.~\ref{fig:compression}(b) often indicate non-nested models, such as the wSBM, as being the best-performing model. Strictly speaking, this is impossible. Non-nested models can always be interpreted as special cases of nested models in which a single hierarchical level is present~\cite{peixoto2014hierarchical}. Since nested models encompass their non-nested variants as special cases, they should always perform at least as well as non-nested models. When indicating non-nested models as being the best-performing ones in Fig.~\ref{fig:compression}(a) and Fig.~\ref{fig:compression}(b), we mean that their corresponding nested variants performed no better, so that the same level of compression can be achieved without the need to consider a nested model. A similar picture holds when the ER model is displayed as optimal, as ER random graph models can be viewed as special cases of the standard SBM in which a single group is present.

A notable feature that can be observed in Fig.~\ref{fig:compression}(a) and Fig.~\ref{fig:compression}(b) is that, for a considerable fraction of CBSAs, the greedy algorithm outperforms the SBM models. It is then interesting to establish how much worse the GA behaves in cases where it is not the best-performing model. Fig.~\ref{fig:compression}(c) displays the description length ratio between the greedy algorithm and the best-performing SBM across all of the metro regions. While SBM models perform better across the entire dataset, the greedy algorithm can consistently achieve compression levels comparable with those of the best-performing SBMs. This casts the greedy algorithm as a suitable proxy to infer near-optimal solutions when contiguity of the inferred clusters is required and reinforces the idea that contiguous spatial regions can adequately capture much of the underlying regularities in the commuting data.

We also assess the number of groups that the proposed methods infer. Fig.~\ref{fig:compression}(d) shows the number of inferred groups as a function of the number of nodes $N$ for the four best-performing SBM models and the proposed greedy algorithm. The first clear result is that nested methods infer a far larger number of groups. This is a well-known property of nested models~\cite{peixoto2017nonparametric, peixoto2014hierarchical}. It can be shown that non-nested SBM models suffer from a \emph{resolution limit} in which the maximum number of groups $B_{max}$ that can be inferred scales as $B_{max} \sim \sqrt{N}$~\cite{peixoto2013parsimonious}. Nested models are able to extend this limit so that $B_{max}$ scales as $B_{max} \sim N / \log N$, which is significantly larger than the resolution limit for non-nested models. This ability to uncover a larger number of groups is also the underlying reason that nested models are able to outperform all other methods, as they can leverage the finer-grained structure in the network to compress it more efficiently. The flip side of this increased expressiveness in the model is that it might be hard to interpret regionalizations with a large number of groups. The nestedness of these models does provide assistance, as it is easier to start from the topmost partition, which displays a more straightforward and coarse-grained interpretation of the underlying structure, and then work our way down the hierarchy. In the case of the regionalization endeavor we are considering, one could, in principle, obtain an informative partition with a lower number of groups by simply regionalizing according to one of the higher levels in the hierarchy. However, the choice of which layer to use for the regionalization is left to the practitioner. Furthermore, the groups inferred by the nested SBMs can present considerable discontiguities in the inferred clusters, even at the coarsest levels of the hierarchy.

Our greedy algorithm infers the fewest groups out of all the considered methods, which is likely due to the imposed contiguity constraints. By only considering partitions that give rise to contiguous spatial units, we limit the possible number of groups to which a node can be assigned. In contrast, the far less constrained non-spatial ensemble of networks that SBM optimization with MCMC explores allows for greater flexibility in the inferred group structures at the expense of sacrificing contiguity.
Nevertheless, the greedy algorithm can still recover a number of groups roughly in line with those inferred by the wSBM (see Fig.~\ref{fig:compression}(e)). Therefore, while not achieving the levels of expressivity that nested models exhibit, we can still retain spatial contiguity without the risk of massively underfitting our data.

Finally, we investigate the similarity of the partitions inferred by the greedy algorithm with those inferred by the SBM models. Fig.~\ref{fig:compression}(f) shows the adjusted mutual information (AMI) \cite{vinh2009information} score between the partitions inferred by the best-performing SBM model with those inferred by the greedy algorithm, along with the difference in description length per node between the two models %
\footnote{
    When comparing against nested models, the partition at the lowest level of the hierarchy was considered, corresponding to the finest-grained resolution.
}.
We can observe that for CSAs, and in particular for entire states, the greedy algorithm is able to infer partitions that have a high degree of similarity with those inferred by the best-performing SBM model. On the other hand, CBSAs exhibit a much larger spread in AMI scores. This spread can be explained in terms of the sizes of the networks considered. The majority of CBSAs displaying very low AMI scores are described by networks comprising a very small number of nodes, often less than $20$. As a result, they are highly susceptible to statistical fluctuations, and any random partition will generally be sufficient to guarantee high enough mutual information scores. Therefore, when the AMI adjusts for random chance, the scores of these partitions decrease dramatically. Finally, if we look at the marginal distribution of AMI scores in Fig.~\ref{fig:compression}(f), we notice the presence of a pronounced spike at an AMI score of one and another smaller one at an AMI of zero. The first spike corresponds to all those metro regions that were grouped into a single cluster by both algorithms. The second corresponds to networks for whom the partitioning, according to the SBM, is extremely similar to a random graph but that were not aggregated into a single cluster by the greedy algorithm.

\subsubsection{Assesment of contiguity violations}

\begin{figure*}[!ht]
    \centering
    \includegraphics[width=\textwidth]{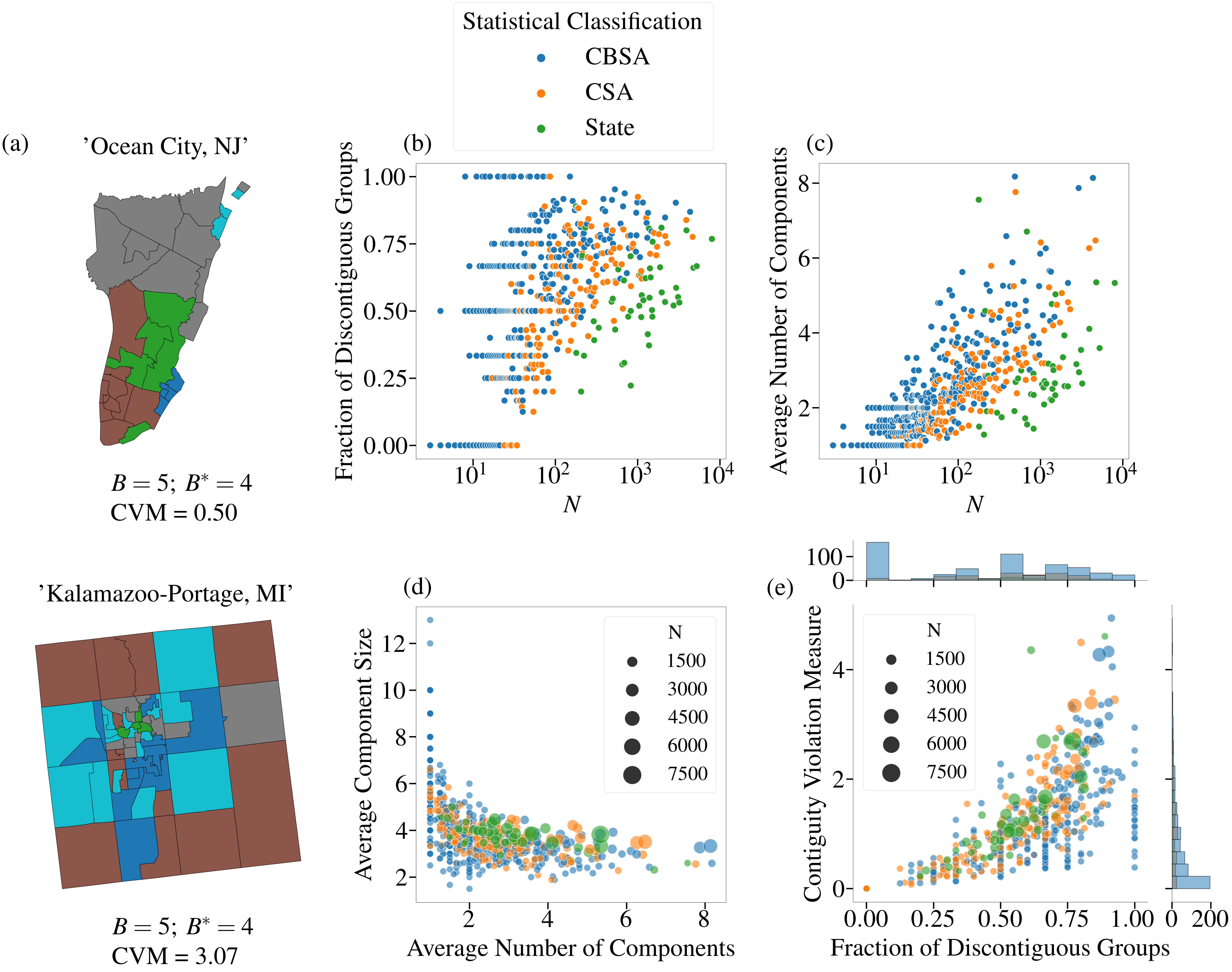}
    \caption{
    (a) Sample regionalizations inferred via the wSBM for Ocean City, NJ, and Kalamazoo-Portage, MI. In both cases, metropolitan areas are divided into $B = 5$ groups, $B^* = 4$ of which are discontiguous. (b) The fraction of discontiguous groups as a function of the number of nodes $N$ across all the metropolitan areas. (c) The average number of components comprising each group as a function of $N$ across all the metropolitan areas. (d) The average component size as a function of the average number of components making up each group across all metropolitan areas. (e) The contiguity violation measure (CVM) (Eq.~\ref{eq:cvm}) as a function of the fraction of discontiguous groups across all the metropolitan areas. The sizes of the points are proportional to $N$ in panels (d) and (e).
    }
    \label{fig:contiguity}
\end{figure*}

As mentioned in Section~\ref{sec:intro}, one of the principal limitations of using SBM models to perform spatial regionalization is that they are not explicitly designed to work with spatial data, so nodes corresponding to regions that might be separated by large spatial distances can be placed within the same group. In turn, this causes the inferred spatial partitions to break up into multiple components, violating the spatial contiguity that is often a desired property of any regionalization endeavor. Fig.~\ref{fig:contiguity}(a) shows two example partitions inferred by a wSBM for the Ocean City and Kalamazoo-Portage CBSAs, each having $B=5$ groups, of which $B^\ast=4$ are discontiguous in space. In what follows, we assess the suitability of SBM models for spatial regionalization tasks by analyzing the degree to which the inferred partitions violate spatial contiguity.

Fig.~\ref{fig:contiguity}(b) shows the fraction of discontiguous groups inferred by the best-performing SBM model across all the metro regions as a function of the number of nodes $N$ in the network. We can observe a considerable spread in the distribution of the fraction of groups that split into multiple components, with no clear dependence on $N$ or the statistical classification of the metropolitan region. For larger networks, SBM models break roughly between $25\%$ to $80\%$ of the groups they infer. The average number of components that make up the inferred clusters is shown in Fig.~\ref{fig:contiguity}(c). In this case, we observe a weak tendency for the number of components that make up each discontiguous partition to increase with $N$, suggesting that the inferred groups tend to be more badly broken as the networks become larger. However, at any fixed $N$, there is again a considerable spread in the average number of components that comprise each group. In contrast, the average size of the components (i.e. the number of tracts making up each component) appears to remain relatively stable regardless of the average number of components that make up each group; see Fig.~\ref{fig:contiguity}(d). The largest variability in the average component size is observed when the average number of components is equal to one, which corresponds to those cases in which the inferred partitions are contiguous. These are typically small networks, with no more than $N = 35$ nodes, and for whom the SBM models infer a maximum number of $B_{wSBM} = 7$ groups. Interestingly, the highest average component size across all metro regions fitted with weighted SBM models is $13$, indicating that, on average, no single component is ever assigned a large number of tracts.

When evaluating regionalization performance, solely considering the fraction of discontiguous groups can ignore valuable information about the characteristics of the discontiguities themselves. For instance, in Fig.~\ref{fig:contiguity}(a), the two samples show that, in both cases, the wSBM identified $B=5$ groups, with $B^*=4$ of them broken up into multiple components. However, a simple visual inspection shows that the nature of the discontiguities in the two cases is not the same. Specifically, the Ocean City CBSA presents a more spatially coherent picture compared to the Kalamazoo-Portage CBSA. In the Ocean City CBSA, all the inferred groups consist of no more than two components, as opposed to the Kalamazoo-Portage CBSA where each group fractures into a larger number of components. Furthermore, the discontinuous groups in the Ocean City case present a more spatially contiguous structure, with a large cluster of contiguous tracts and an additional lone tract separated from the main cluster (see, for example, the green or gray-colored groups). This allows for easier identification of dominant clusters and more apparent spatial patterns. In contrast, the clusters within the Kalamazoo-Portage CBSA split into multiple roughly equally sized components that can lie at considerable spatial distances from each other, thereby making the interpretation of the regionalization results harder.

To capture these effects and distinguish among different ways in which discontiguities occur, we introduce a contiguity violation measure (CVM) as follows. Suppose we have a partition of a network of $N$ nodes into clusters, of sizes $\{n_r\}$ as before. Now suppose that $n_{rc}$ nodes fall in the $c$-th connected component of cluster $r$, and there are $C_r$ connected components in total corresponding to cluster $r$. We want a measure that penalizes greater $C_r$---more components for cluster $r$---but that provides further nuance by only slightly penalizing components $c$ with small sizes $n_{rc}$ relative to the total cluster size $n_r$. One can adapt the ``effective number of groups'' measure from the community detection literature (used, for example, in \cite{riolo2017efficient}) to simultaneously capture both of these effects. The effective number of components $C_{\text{eff}}(r)$ for cluster $r$ is then given by
\begin{align}\label{eq:Ceff}
C_{\text{eff}}(r) = \exp \left( -\sum_{c=1}^{C_r}\frac{n_{rc}}{n_r}\ln \left(\frac{n_{rc}}{n_r}\right) \right).     
\end{align}
Eq.~\ref{eq:Ceff} will achieve a maximum value of $C_r$ when all of cluster $r$'s connected components $c$ are of equal sizes $n_{rc}=n_r/C_r$, and will take on a value close to $1$ when only a single connected component contains most of the nodes in cluster $r$.

Given Eq.~\ref{eq:Ceff}, we define the contiguity violation measure as
\begin{equation}\label{eq:cvm}
    \text{CVM} = \frac{\sum_{r=1}^B C_{eff}(r) - B}{B},
\end{equation}
where $B$ is the total number of inferred groups. We can interpret Eq.~\ref{eq:cvm} as telling us the typical number of additional connected components there are per community in the network. In the perfect case in which no group is broken, $\sum_r C_{eff}(r)=\sum_r(1)=B$ and $\text{CVM}=0$, as there are no excess components in any community. On the other hand, if the groups split into clusters, then $\sum_r C_{eff}(r) > B$ and the CVM increases. In the most extreme case where all clusters $r$ break up into $n_r$ components of spatially isolated nodes, we have $\text{CVM}=\frac{N}{B}-1$. The CVM allows us to capture both desired mechanisms of discontiguity, as it will increase as the number of components per group increases but will also penalize the discontiguities differently according to the distribution of component sizes.

Fig.~\ref{fig:contiguity}(e) shows the value of the CVM along with the fraction of discontiguous groups across all metro regions considered in our study. We can notice that metropolitan areas presenting a small fraction of discontiguous groups tend, as expected, to have low values of the CVM. As the fraction of broken groups increases, the CVM can take on a wide range of values, depending on the nature of the discontiguity. If we consider all states, the average CVM is given by $1.37 \pm 0.92$, corresponding to about $1.45$ to $3.29$ effective components per group.

Fig.~\ref{fig:ny_sample} in Appendix~\ref{appendix:appendix_b} shows additional sample partitions obtained by the wSBM and the greedy agglomerative method for the New York-Newark-Jersey City CSA, along with the CVM for the wSBM.

\subsubsection{Assesment of existing administrative boundaries}

\begin{figure*}
    \centering
    \includegraphics[width=\textwidth]{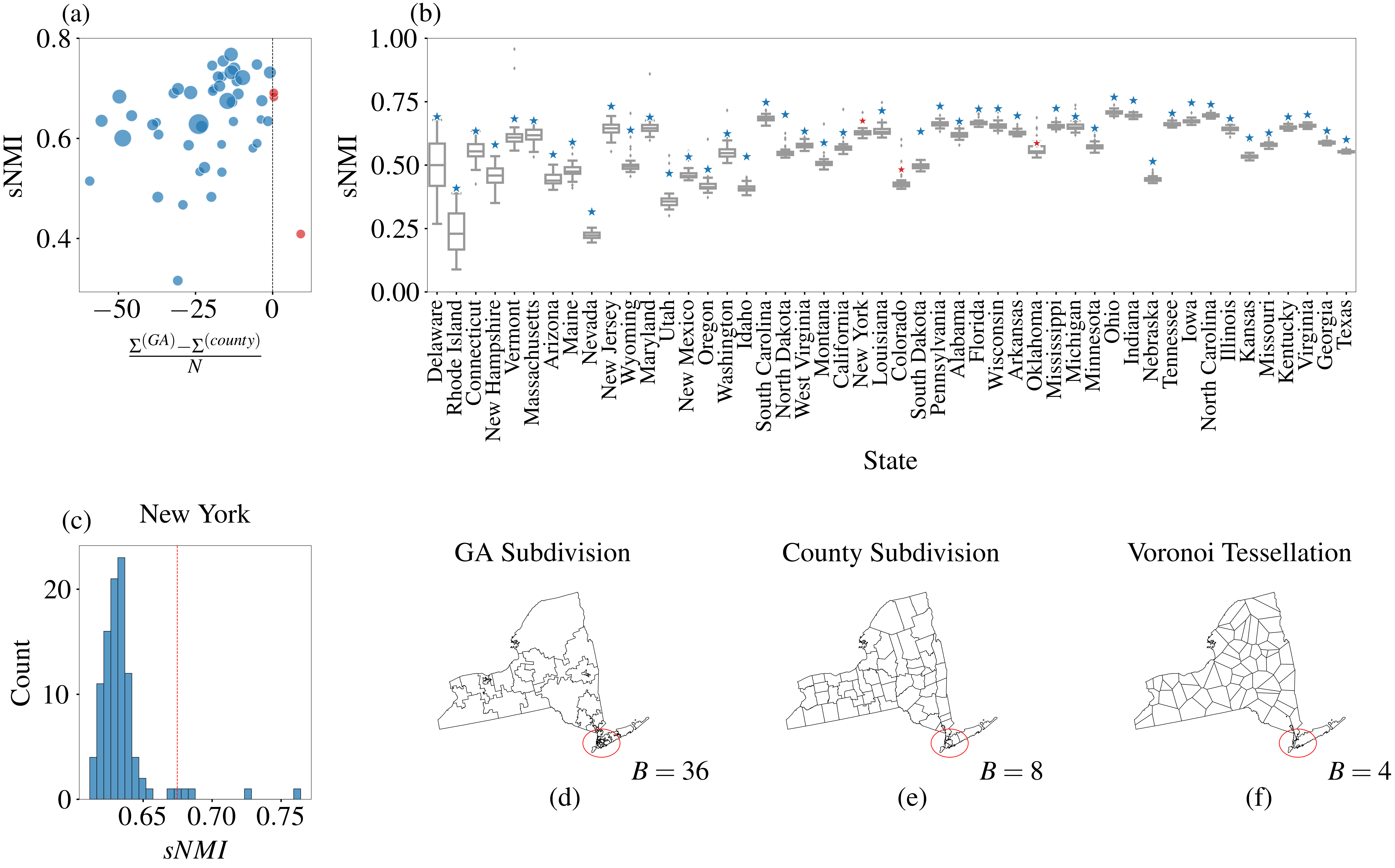}
    \caption{
    (a) The spatial normalized mutual information (sNMI) score (Eq.~\ref{eq:snmi}) between the partitions inferred via the greedy agglomerative algorithm and those induced by the administrative county subdivisions, for all the lower 48 states in the U.S.A., as a function of the difference in description length $\Sigma^{(GA)} - \Sigma^{(county)}$ normalized by the number of nodes $N$. The sizes of the points is proportional to $N$. For all states, the agglomerative was stopped when the number of inferred groups $B_{GA}$ was equal to the number of counties in each state, $B_{GA} = B_{county}$, for a more direct comparison of their partitions. Red points indicate states where the county subdivision has a lower description length than the one inferred by the greedy algorithm due to the artificial constraint $B_{GA} = B_{county}$ which may lie far from the true optimal $B$. (b) Boxplots of the distributions of sNMI scores between the partitions inferred by the greedy algorithm and $100$ random Voronoi tessellations with a number of seeds equal to $B_{GA}=B_{county}$. The original sNMI scores between the greedy solution and the administrative county subdivisions are indicated by a star; blue stars indicate county partitions whose sNMI scores with the corresponding greedy solution lie above the 95th percentile of the boxplot distributions, and red stars indicate solutions that lie below. (c) Distribution of sNMI scores between the partition inferred via the greedy algorithm (GA) and $100$ random Voronoi tessellations for the state of New York. The vertical dashed red line indicates the sNMI score between the original administrative county subdivision and the GA partition. (d) Regionalization results for the state of New York inferred via the greedy algorithm. (e) Administrative county subdivision of New York state. (f) A sample Voronoi tessellation of the state of New York. In all cases, red circles indicate the number of regions in the New York City area. As can be observed, the greedy algorithm uses the majority of the inferred groups to describe this region, as would be expected based on its high level of commuting heterogeneity and high tract density.
    }
    \label{fig:admin_bounds}
\end{figure*}

We now turn our attention to analyzing how well the current administrative boundaries subdividing U.S. states into counties capture structural regularities in commuting patterns and, therefore, faithfully reflect the underlying labor market structure.

To perform this assessment, we can compare the tract partitions correpsonding to the county subdivisions with the results of applying the greedy agglomerative algorithm to the commuting flows across the state. This allows us to assess the extent to which county subdivisions compress the commuting data, and to what extent these subdivisions are close to the information theoretically optimal coarse-graining of the tracts according to commute structure. Specifically, we consider the case in which we enforce the number of groups our algorithm infers, $B_{GA}$, to equal the number of counties, $B_{county}$, in each state. This reflects the common scenario in spatial regionalization where it is necessary for the practitioner to externally enforce the desired number of group subdivisions.

When comparing spatially contiguous clusterings, it is important to be cautious of the effects that the contiguity constraint might have on measures of similarity, such as mutual information. By enforcing contiguity we significantly limit the number of valid subdivisions we can obtain, and as a result two subdivisions that are uncorrelated apart from both being spatially contiguous may appear similar because standard mutual information-based measures are assessing their similarity relative to \emph{all} partitions (most of which are discontiguous) \cite{kirkley2022spatial,kirkley2024dynamic}. To circumvent this limitation, we introduce a spatial variant of the normalized mutual information (sNMI), which allows us to compare the similarity scores of the two partitions with those obtained by a suitably chosen null model for random spatially contiguous clustering. We first introduce the sNMI measure and then discuss how it can be used to compare similarity scores with those achieved by a suitably chosen null model. 

Given a state of interest, let $r\in \{1,...,B_{county}\}$ be counties and $s\in \{1,...,B_{GA}=B_{county}\}$ be the spatial communties inferred by the greedy agglomerative method for this state. Also let $a(u)$ be the area of a spatial unit $u$ under consideration such that $a(r)$ is the area of county $r$ and $a(s)$ is the area of cluster $s$ in the greedy partition. Denote the total area of the state of interest as $a_{state}$, which satisfies $a_{state}=\sum_{r}a(r)=\sum_{s}a(s)$. The mutual information between the county and greedy partitions would normally be calculated by computing the entropy of the group size distributions $\bm{n}/N$ within each partition then subtracting the entropy of the group sizes in the joint partition defined by the intersections of all pairs of groups in the two partitions \cite{vinh2009information}. This tells us the difference in description length it takes to transmit both partitions separately versus transmitting them by first transmitting the intersections of their clusters. To capture the actual spatial similarity among the partitions rather than their node-level similarity, one can extend this concept to consider the distributions of spatial sizes among the clusters within the two partitions separately and their intersection partition.

Mathematically, if we select a point in space---crucially, not necessarily a network node---uniformly at random from the state of interest, the probability that it falls in county $r$ is given by $a(r)/a_{state}$. Similarly, the probability that a randomly selected point in space falls into cluster $s$ in the greedy partition is $a(s)/a_{state}$. The probability that the point falls into \emph{both} $r$ and $s$ is given by $a(r\cap s)/a_{state}$. The three distributions $\{a(r)/a_{state}\}$, $\{a(s)/a_{state}\}$, and $\{a(r\cap s)/a_{state}\}$ can then be used to compute the two partition entropies and join partition entropy respectively, whose difference is the mutual information. We can then define a spatial version of the NMI, which we call the sNMI, as
\begin{equation}\label{eq:snmi}
    \text{sNMI} = \frac{H_{county} + H_{GA} - H_{county,GA}}{\frac{1}{2}\left[ H_{county} + H_{GA}\right]},
\end{equation}
where 
\begin{align}
H_{county} &= -\sum_{r}\frac{a(r)}{a_{state}}\log \left(\frac{a(r)}{a_{state}}\right), \\ 
H_{GA} &= -\sum_{s}\frac{a(s)}{a_{state}}\log \left(\frac{a(s)}{a_{state}}\right) 
\end{align}
are the spatial entropies of the county and greedy partitions respectively, and
\begin{align}
H_{county,GA} = -\sum_{r,s}\frac{a(r\cap s)}{a_{state}}\log \left(\frac{a(r\cap s)}{a_{state}}\right)    
\end{align}
is the spatial entropy of their intersection. The sNMI will be equal to $0$ when the county and greedy partitions are completely spatially uncorrelated, and equal to $1$ if and only if they are perfectly equal. 

Fig.~\ref{fig:admin_bounds}(a) shows the sNMI scores between the county subdivisions and the partitions inferred by the greedy algorithm for all the lower 48 states in the U.S., along with the difference in description length per node $(\Sigma^{(GA)} - \Sigma^{(county)}) / N$, where $\Sigma^{(county)}$ has been evaluated with Eq.~\ref{eq:Wdl}. The most striking feature we can observe is that for three of the states (shown in red in Fig.~\ref{fig:admin_bounds}(a)), the county subdivisions can actually outperform the greedy algorithm in terms of compression. This is because, by stopping the algorithm when $B_{GA} = B_{county}$, we are interrupting the regionalization procedure either before it has reached the minimum or after it has overshot it (See Fig.~\ref{fig:methodology}.) In all other cases, the greedy algorithm outperforms the administrative county subdivisions in terms of description length, usually by a significant margin.

Almost all states display high sNMI scores between the county subdivision and the partitions inferred by the greedy algorithm---see Fig.~\ref{fig:admin_bounds}(a). However, as stated previously, one must be careful to interpret these results, as they might be artificially inflated by contiguity constraints. To control for this, we randomly partition each state by randomly selecting $B_{county}$ points (``seeds'') within the state and then generating a Voronoi tessellation starting from these seeds. A Voronoi tessellation clusters the space by assigning each point in space to the cluster of its closest seed in terms of Euclidean distance. Fig.~\ref{fig:admin_bounds}(f) shows a sample tessellation for the state of New York. 

Fig.~\ref{fig:admin_bounds}(b) shows the distribution of sNMI scores between the partitions inferred by the greedy algorithm and $100$ different random tessellations for each state. The stars in the plot indicated the original sNMI score between the greedy partition and the county subdivision. If this value lies above the 95th percentile of the random Voronoi distribution, the star is colored blue. Otherwise, it is colored red. A more detailed view of the sNMI distribution is shown in Fig.~\ref{fig:admin_bounds}(c) for the case of New York. Fig.~\ref{fig:admin_bounds}(b) displays the confounding effect that spatial contiguity can have in attempting to assess the cluster similarity, as also random but spatially contiguous partitions are sufficient to obtain considerably high sNMI scores. However, in most cases, the county subdivisions appear to perform significantly better than random ones, indicating that they are truly capturing some level of structural regularity in the commuting networks.

Finally, in Fig.~\ref{fig:admin_bounds}(d)-(f), we show a sample of the regionalizations used in our analysis for the case of New York state. Fig.~\ref{fig:admin_bounds}(d) displays the group structure inferred by our greedy algorithm when $B_{GA} = B_{county}$, Fig.~\ref{fig:admin_bounds}(e) displays the county subdivision, and Fig.~\ref{fig:admin_bounds}(f) shows a sample random Voronoi tessellation of the state of New York with a number of seeds equal to $B_{county}$. The principal difference between the greedy partition and the other two subdivisions lies in the way the clusters are distributed in space. Although all three partitions have the same number of groups, the county and Voronoi partitions tend to have equally sized groups, while the partition inferred by the greedy algorithm places over half of the groups around a small spatial region corresponding to the New York City area. This is where the majority of the commute flows are located and commuting patterns become increasingly complex, so it is information theoretically more justified to delineate a larger nunber of clusters in this area to capture these heterogeneities. This pattern of increasing the number of units around key urban areas so as to better capture the complexity of commute flows while ascribing fewer and larger groups to capture more peripheral areas is common across all of the states considered in the study.

\subsection{County-level results}

\begin{figure*}
    \centering
    \includegraphics[width=\textwidth]{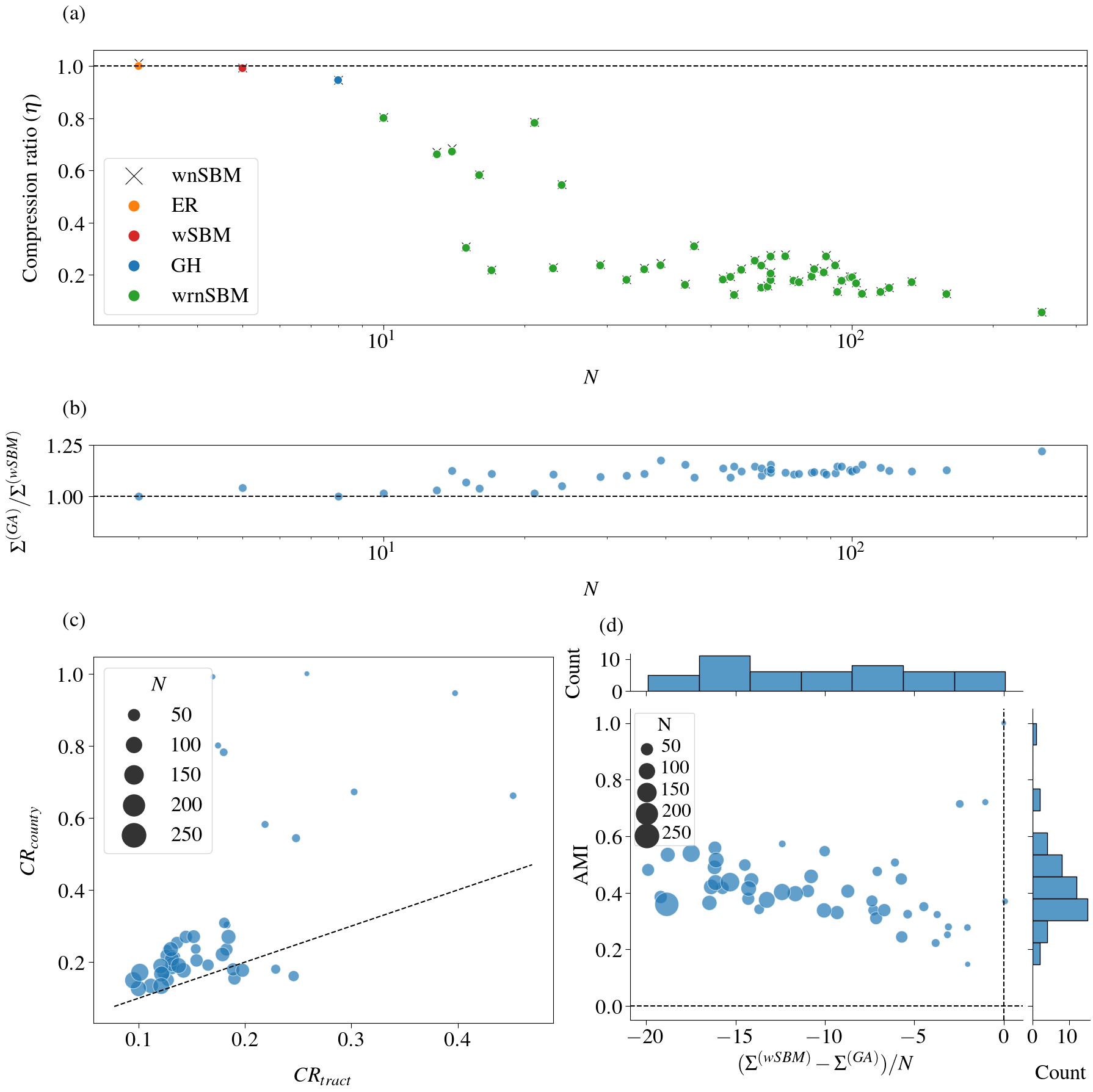}
    \caption{
        (a) Compression ratio ($\eta$) for the best-performing wSBM model when the commuting flows are aggregated at the county level, for all lower 48 states. Black crosses indicate the CR values for the wnSBM model, which was dominant at the tract scale (see Fig.~\ref{fig:compression}). Although the wnSBM is never the optimal model when counties are the fundamental areal unit, it still achieves near-optimal compression levels. (b) Description length ratio between the regionalizations obtained by the greedy algorithm the best-performing wSBM, for each state. As is the case at the tract level, while SBM models outperform the greedy algorithm across all states, it is still possible to achieve comparable levels of compression via the greedy algorithm. (c) Compression ratios obtained by the best-performing wSBM model at the tract and county levels when partitioning each state. The black dashed line indicates the case where the county-level CR, $CR_{county}$, equals the tract-level CR, $CR_{county} = CR_{tract}$. (d) The adjusted mutual information score (AMI) between the partitions inferred by the best-performing wSBM model and the greedy algorithm as a function of the description length difference $\Sigma^{(wSBM)} - \Sigma^{(GA)}$ divided by the number of nodes $N$. The sizes of the points are proportional to $N$.
    }
    \label{fig:county_level_compression}
\end{figure*}

\begin{figure*}
    \centering
    \includegraphics[width=\textwidth]{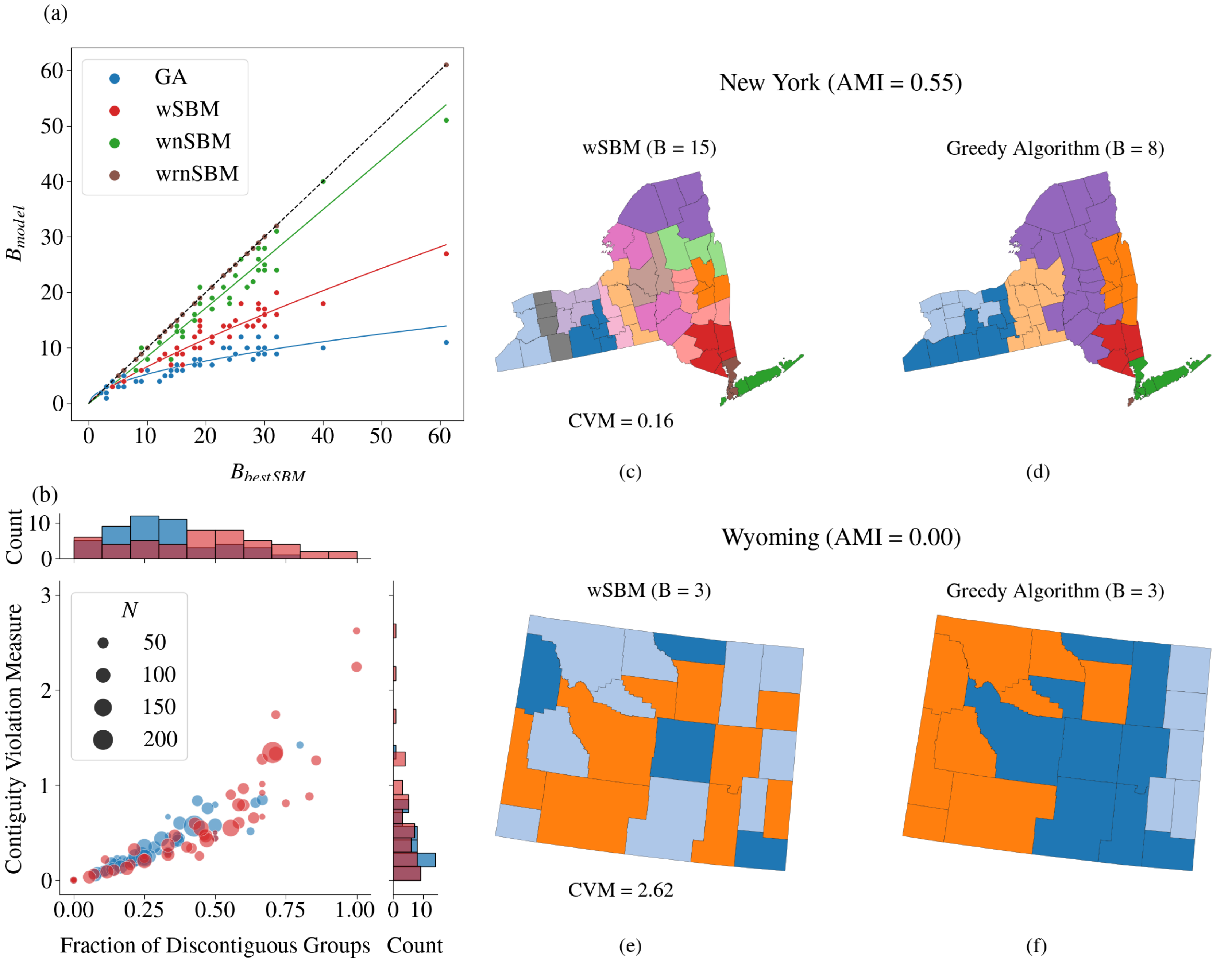}
    \caption{
        (a) Number of groups inferred by different models versus the number of groups inferred by the best-performing wSBM model, $B_{bestSBM}$, across the lower 48 states when flows are aggregated at the county level. The dashed black line indicates the case $B_{model} = B_{bestSBM}$. (b) Contiguity violation measure (CVM) (Eq.~\ref{eq:cvm}) as a function of the fraction of discontiguous groups for the best-performing wSBM model (blue points) and for the standard wSBM (red points). Point sizes are proportional to $N$. (c) Regionalization inferred via the wSBM for the state of New York. (d) Regionalization inferred via the greedy algorithm for the state of New York. (e) Regionalization inferred via the wSBM for the state of Wyoming. (f) Regionalization inferred via the greedy algorithm for the state of Wyoming. In both cases the contiguity violation measure is shown for the models inferred via the wSBM.
    }
    \label{fig:county_level_contiguity}
\end{figure*}

In this section, we address the question of how the aggregation scale at which the spatial data is considered can impact the effectiveness of the previously studied methods as tools to perform spatial regionalization. As discussed in Section~\ref{sec:results}, census tracts have the desirable property of being part of a well-defined geographical hierarchy, meaning that we can aggregate tracts to obtain counties, which can themselves be aggregated into entire states. (In fact, this is done in \cite{he2020demarcating} for regionalization.) In what follows, we consider the results of performing spatial regionalization across all lower 48 states in the U.S. when counties are used as the fundamental unit of spatial aggregation instead of tracts.

To construct commuting networks at the county scale, we combined all tracts into their respective counties, which we then mapped to the nodes of a weighted directed network. Similarly, we combined the tract-to-tract flows to compute the commute flows between counties and mapped them to the edges of the network. Finally, we ran all the previously considered weighted variants of the SBM models and our greedy method on these networks.

Fig.~\ref{fig:county_level_compression}(a) shows the compression ratio achieved by the best-performing SBM as a function of the number of nodes $N$ in the network for all lower 48 states. One can observe that, even at the county scale of spatial aggregation, SBM models can still achieve good levels of compression compared to the na\"ive transmission of the ER null model. Notable exceptions happen for the smallest states, where, in at least one case, no model is able to perform better than the ER random graph. 
Similarly, most states with compression ratios above $0.5$ have fewer than $20$ nodes. It is perhaps more surprising that good levels of compression can still be achieved for a number of states of this size, indicating that, even at the county level of aggregation, block structures are still able to efficiently capture the underlying structure of commute flows.

Another novelty is the emergence of ranked models as the best-performing ones, in contrast to the previously seen tract-level results in which non-ranked nested models were dominating. It is possible that by broadening the scale of spatial aggregation a more exaggerated flow hierarchy between commute regions emerges. However, if we plot the CR values obtained by the wnSBM, shown as black crosses in Fig.~\ref{fig:county_level_compression}(a), we can notice that they achieve almost exactly the same CR values as the best-performing models across the entire dataset, suggesting that, while a ranking might emerge, its ability to explain the data is still relatively weak compared to the nestedness of the groups.

In Fig.~\ref{fig:county_level_compression}(b), we show the description length ratio between the partitions obtained via our greedy algorithm and those inferred by the best-performing SBM. As was the case at the tract level, we observe that SBM models outperform the greedy method across the entire dataset. However, excellent levels of compression can be achieved by the greedy method without the need to sacrifice contiguity, indicating that spatially contiguous regions can suitably capture the underlying commute structure at the county scale as well.

While all models considered can achieve good levels of compression even at the county level, they can generally achieve better CR values when census tracts are considered as the fundamental spatial unit---see Fig.~\ref{fig:county_level_compression}(c). The highest points along the $CR_{county}$ axis (representing the compression ratio values at the county level) can, to a reasonable degree, be explained in light of the fact that they correspond to networks comprised of a very small number of nodes. The rest of the points still tend to fall above the $CR_{county} = CR_{tract}$ line, indicating that our models can generally achieve a higher compression at the tract level than at the county level. However, it must be noted that there is a considerable disparity in the network sizes at these two different resolutions. While it is true that counties are composed of aggregates of census tracts, the difference in resolution can sometimes become quite staggering. For example, California, the metro region with the highest number of tracts ($7974$), only has $58$ counties. In turn, it becomes much easier to achieve good levels of compression with respect to the na\"ive case when considering U.S. states at the tract level. Somewhat surprisingly, this is not the case for all states, and lower CR values can still be achieved at the more coarse-grained county level scale, indicating that broadening the resolution scale can, in some cases, aid in inferring commuting patterns that might be obscured by noise at finer resolutions.

In Fig.~\ref{fig:county_level_compression}(d), we plot the AMI scores between the partitions inferred by the best-performing SBM model and those inferred via our greedy method as a function of the difference in description length per node between the two models. Curiously, the AMI scores have dropped with respect to the tract level case, indicating that the two methods are finding more dissimilar partitions. Examples of the partitions inferred by the wSBM and our greedy method are shown in Fig.~\ref{fig:county_level_contiguity}(c)-(f).

We also compare the number of groups inferred by the various methods. Fig.~\ref{fig:county_level_contiguity}(a) shows the number of groups inferred by four different models (the greedy algorithm, the weighted SBM, the weighted nested SBM, and the weighted ranked nested SBM) with those inferred by the best-performing SBM model. Once more, we observe that nested models can infer a larger number of groups. By looking at Fig.~\ref{fig:county_level_contiguity}(a), we observe that the points corresponding to the wrnSBM lie almost exclusively along the diagonal, as this is the dominating model at the county level. The wnSBM line lies just below it. By contrast, the wSBM and the greedy algorithm infer a considerably lower number of groups, as before. Interestingly, the number of groups inferred does not appear to have a large effect on how discontiguous they are. Fig~\ref{fig:county_level_contiguity}(b) shows the contiguity violation measure (Eq.~\ref{eq:cvm}) as a function of the fraction of discontiguous groups. Blue points indicate CVM scores for the best-performing SBM models, and red points those for the wSBM. While wSBMs do infer partitions with a larger fraction of discontiguous groups for some states, they are mostly in line with what we observe for the best-performing SBM models, and their CVM scores can often be lower.

An important question to address is whether increasing the spatial aggregation scale affects the contiguity of the partitions inferred by SBM models. If we look at the results shown in Fig.~\ref{fig:county_level_contiguity}(b), we observe that even at this more coarse-grained spatial aggregation scale, the partitions inferred by the SBM models can present a high degree of contiguity violation, as shown by the high fraction of discontiguous groups that can be achieved in some cases. However, the overall performance across all states is nevertheless improved. By focusing on the best-performing models and looking at the marginal distribution for the fraction of discontiguous groups, we observe that the mean of the distribution lies around $0.3$. At the tract level, the average fraction of discontiguous groups across all states was about $0.55$. So, while SBM models can still infer heavily discontiguous clusterings, they perform better at coarser aggregation scales. Perhaps more important is the significant decrease in the CVM with respect to what was observed at the tract level in Fig.~\ref{fig:contiguity}(e). For the best-performing SBMs, the average CVM across all states at the county aggregation scale is $0.36 \pm 0.27$, corresponding to an effective number of components per group between $1.09$ and $1.63$, with an average of $1.36$. This is a considerable improvement over the tract-level results, which gave, on average, $2.37$ components per group. This indicates that, although SBM models will in most cases infer partitions that violate spatial contiguity, a single component will be responsible for most of the nodes in the group.

In Figs.~\ref{fig:county_level_contiguity}(c)-(d), we display samples of the partitions inferred via the wSBM and the greedy algorithm for the state of New York. In the case of the wSBM we now have three discontiguous groups. Although one of these groups (shown in pink) breaks into two almost symmetric clusters with two and three nodes respectively, respectively, the other two discontiguous groups fracture in the most heterogeneous way possible. They are composed of a dominating cluster of contiguous counties and a single isolated county spatially removed from the cluster. This allows the partition inferred by the wSBM to have a low CVM score ($0.16$) even though $20\%$ of the inferred clusters are broken. We also notice that the partitions inferred by the two algorithms differ significantly. In this case, as the partition inferred by the wSBM has a low CVM score and a lower description length, it should probably be preferred as a regionalization method. However, depending on the degree of spatial contiguity desired, this might not always be the appropriate choice. In these situations, the CVM could be used as a guideline to aid the decision. For example, Fig.~\ref{fig:county_level_contiguity}(e) and (f) show the regionalization results obtained by the wSBM and the greedy algorithm for the state of Wyoming. In both cases, the models infer three groups. However, the partition inferred by the wSBM is highly discontiguous, and its groups are fractured into approximately equally sized components. Arguably, this is not a satisfactory regionalization of the state, and the CVM score reflects this, indicating that there are on average around $3.6$ effective components per group.

\section{\label{sec:conclusions} Conclusion}

We have systematically assessed the suitability of a wide range of stochastic block models (SBMs) for urban boundary delineation with commuting flows. To provide a practical and efficient solution for explicit regionalization with community detection methods, we have developed a fast greedy agglomerative algorithm that is adaptable to a wide range of objectives including the SBMs of interest. Our results show that, while SBMs are able to achieve significant data compression, they can also present considerable degrees of spatial discontiguity at low spatial aggregation scales, making them mostly unsuitable for spatial regionalization in this regime. Conversely, we have shown that the proposed greedy algorithm can achieve similar levels of compression as those of more expressive SBM models without sacrificing spatial contiguity, making it a more viable option for performing regionalization at low spatial aggregation scales. As the fundamental areal units become more coarse-grained, SBMs can infer partitions with lower contiguity violation scores, potentially making them suitable for regionalization tasks provided the aggregation scale is large enough. Furthermore, we have shown that the chosen network representation can profoundly affect the quality of the inferred regionalization, with weighted models typically achieving far better compression and fewer groups than multigraph representations. We have also analyzed the extent to which current administrative counties capture the underlying commute flow structure when compared with our greedy agglomerative algorithm and found that, while not optimal, they are still able to capture the underlying commute structure in a statistically significant way compared to  a suitable null model. 

The work presented here can be extended in a variety of ways. For example, our greedy algorithm could be adapted to accommodate more expressive objective functions. In the current paper, we have specifically focused on capturing the interaction patterns between nodes but disregarded potentially important information about the spatial units that can be incorporated in the form of node metadata \cite{kirkley2022spatial}. By integrating node-level information with the commuting interactions among spatial units, one could analyze their combined effect on spatial regionalization as well as assess the contributions of each data source to the regionalization procedure. Furthermore, in principle, it is possible to stack multiple types of node metadata and edge interactions within the same framework, provided we can calculate the required combinatorial expressions to compute the description length. This would allow us to model increasingly complex interactions between the single spatial units. Another direction of interest would be to develop hierarchical encodings, which would enable us to capture information across multiple spatial scales in a coherent manner.

Stochastic block models could also be extended to better suit spatial networks, making them more directly applicable to urban boundary delineation tasks. For example, the current formulations for weighted networks aim to be as general as possible. As such, they generally assume maximum entropy distributions over the edge covariates. While this is a very principled approach, it might not be best suited to spatial networks, for which some universal patterns are well known. For example, the gravity law~\cite{zipf1946p, erlander1990gravity} stipulates that the volume of flow between two regions should be proportional to the population sizes and inversely proportional to their distance, and has been shown to provide good fits of urban mobility data~\cite{kwon2023multiple}, with gravity-like models often rivaling the predictions of far more complex (and non-interpretable) machine learning models~\cite{cabanas2023human, ruzicska2024can}. As such, it would be of interest to incorporate information about distances or populations in the priors for the edge covariates. One could also consider more narrowly focused versions of the SBM. As mentioned previously, one of the factors that makes SBM models appealing is their ability to capture generic mesoscale topologies. However, in light of the above results, it might be beneficial to focus specifically on assortative mixing patterns for regionalization endeavors. In~\cite{zhang2020statistical}, Zhang and Peixoto propose an assortative version of the SBM, which could be adapted and applied to perform spatial regionalization as done in this work.

\section{Acknowledgements}
A.K. was supported by the Hong Kong Research Grants Council under ECS--27302523 and an HKU Urban Systems Institute Fellowship.
Grant. The authors thank Baiyue He for participating in data collection and preparation. 




\clearpage

\appendix
\section{\label{appendix:appendix_a} Appendix A: Algorithm time complexity}

\begin{figure}[!ht]
    \centering
    \includegraphics[width=\columnwidth]{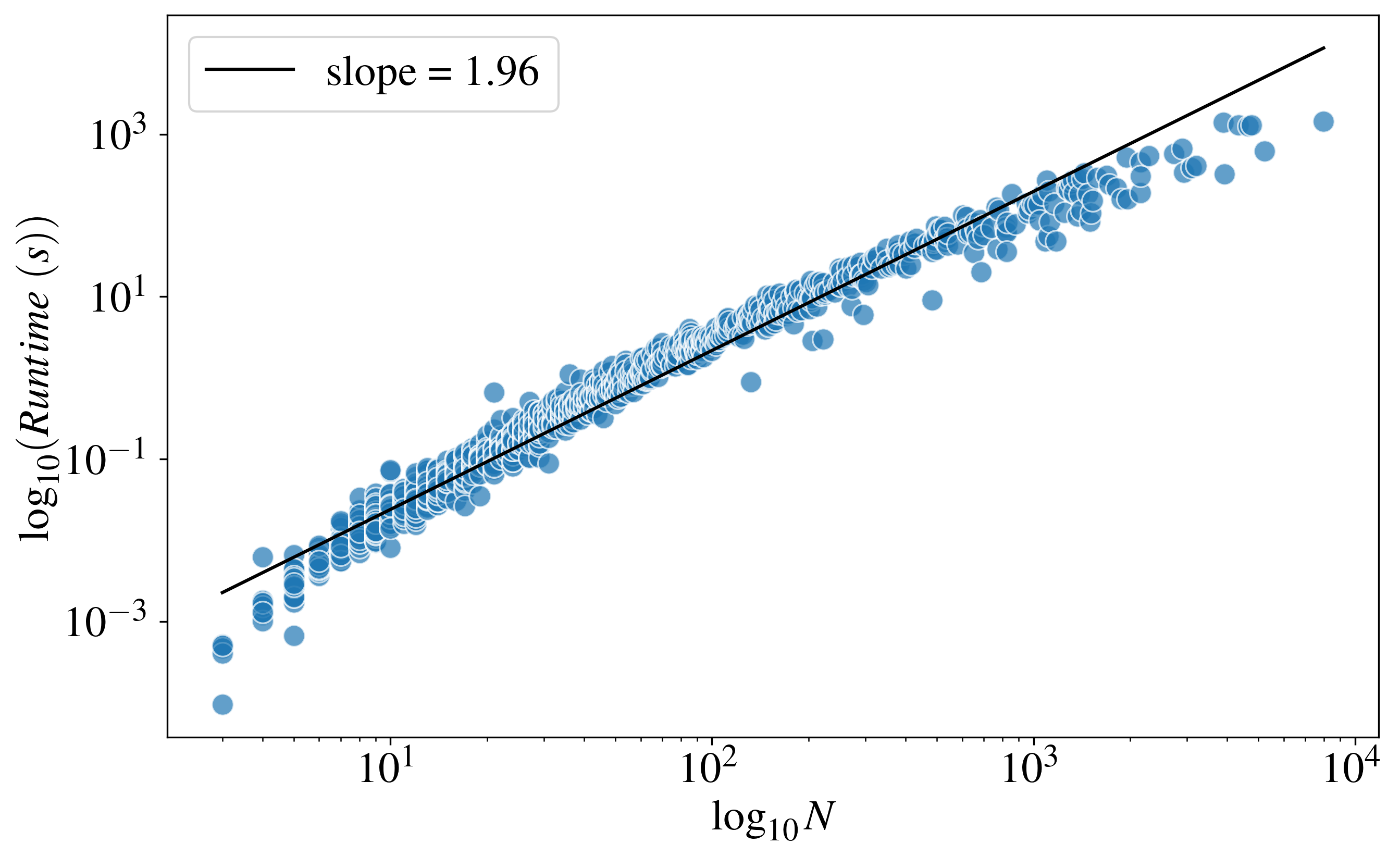}
    \caption{
        Runtimes of the greedy agglomerative algorithm as a function of the number of nodes in the network across all metropolitan areas considered in this study. While its theoretical time complexity is exactly $O(N^2)$ (see Sec.~\ref{sec:greedy}), the results indicate subquadratic scaling of our algorithm in practice.
    }
    \label{fig:exec_time}
\end{figure}

In Fig.~\ref{fig:exec_time}, we plot the run time of our algorithm as a function of the number of nodes $N$ for all the metropolitan areas considered in this study. The relationship exhibits subquadratic scaling in $N$ (see Sec.~\ref{sec:greedy}).

\section{\label{appendix:appendix_b} Appendix B: Example inferred urban boundaries}

\begin{figure*}
    \centering
    \includegraphics[width=\textwidth]{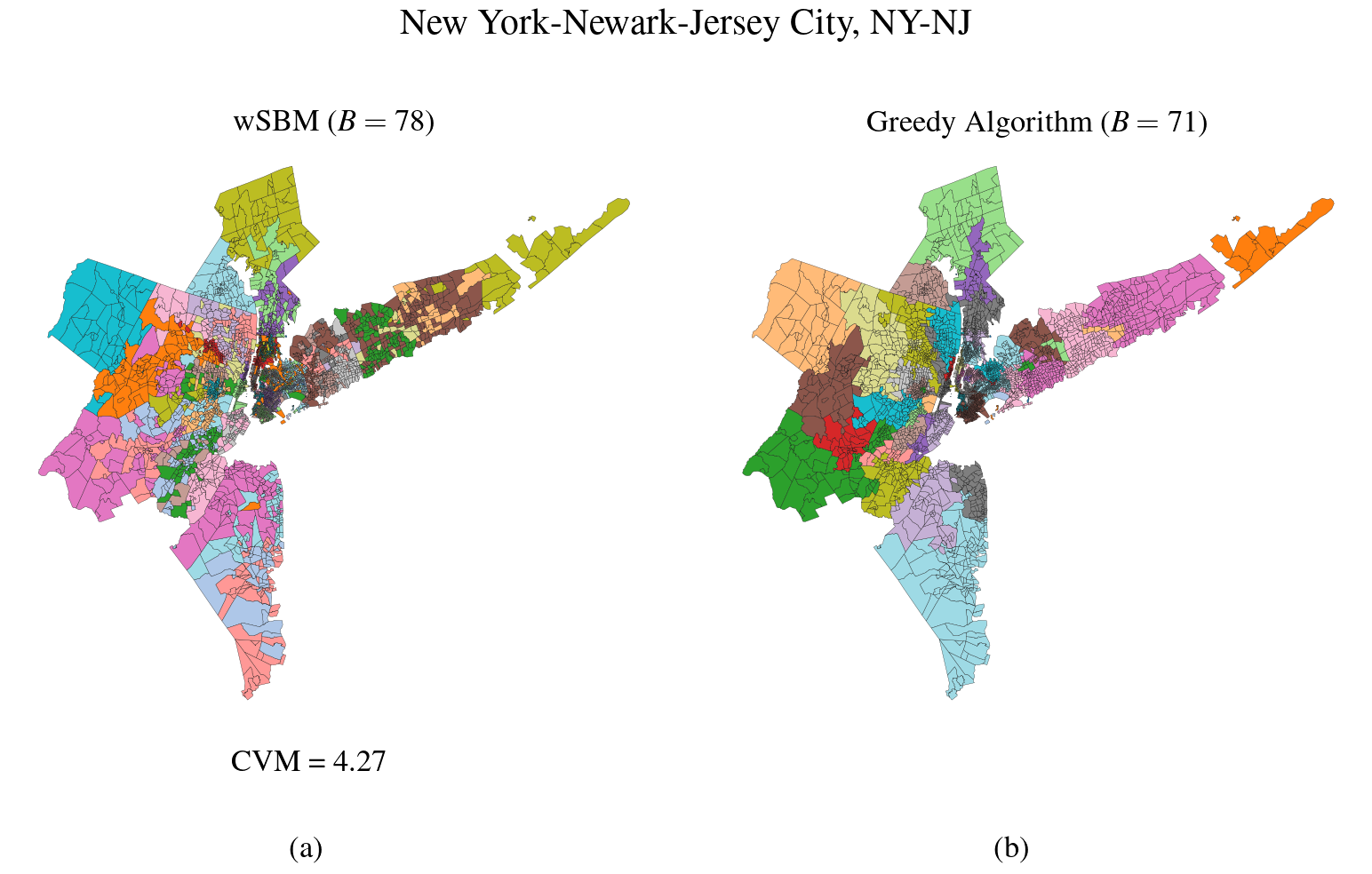}
    \caption{
        Partitions inferred by the best-performing SBM model (a) and the greedy agglomerative algorithm (b) for the New York-Newark-Jersey City, NY-NJ CSA. Although both models infer a similar number of groups and, to a reasonable degree, similar mesoscale structures, the contiguity violation measure for the wSBM is approximately $4.27$, meaning that, on average, each inferred group has $5.27$ effective connected components. High values of the CVM such as this indicate considerable spatial contiguity violations and poor regionalization performance.
    }
    \label{fig:ny_sample}
\end{figure*}

In Fig.~\ref{fig:ny_sample}, we display sample regionalizations obtained by fitting a wSBM (panel a) and our greedy algorithm (panel b) to the New York-Newark-Jersey City CSA. We notice that both methods infer a similar number of groups ($78$ for the wSBM and $71$ for the greedy method) and capture similar mesoscale structures to a large degree. However, the wSBM infers heavily discontiguous groups, as indicated by a high value of the CVM corresponding to roughly $5.3$ effective connected components per inferred group.

\end{document}